\documentclass[aps,prd,onecolumn,showpacs,nofootinbib,amsmath,amssymb,floatfix,superscriptaddress,showkeys]{revtex4}

\def\Nature{{ Nature~}}
\def\Science{{ Science~}}

\def\PRD{{ Phys. Rev. D~}}
\def\PRC{{ Phys. Rev. C~}}
\def\PRL{{ Phys. Rev. Lett~}}
\def\ApJ{Astrophys. J.~}
\def\ApJL{Astrophys. J. Lett.~}
\def\AAA{Astron. Astrophys.~}
\def\MNRAS{Mon. Not. R. Astron. Soc.~}
\def\ARAA{Ann. Rev. Astron. Astrophys.~}
\def\PhRvD{Phys. Rev. D~}

\def\etal{{\it et al.~}}
\def\be{\begin{equation}}
\def\ee{\end{equation}}
\def\bea{\begin{eqnarray}}
\def\eea{\end{eqnarray}}

\usepackage{longtable}
\usepackage{booktabs}

\usepackage{graphicx}
\usepackage{dcolumn}
\usepackage{bm}
\usepackage{captcont}
\usepackage{xcolor}
\usepackage{threeparttable}

\usepackage{epstopdf}
\usepackage{mathtools}
\usepackage{natbib}
\usepackage{ulem}
\usepackage{color,txfonts}
\definecolor{blue0}{rgb}{0,0,0.6}
\usepackage[colorlinks,linkcolor=blue0,anchorcolor=blue0,citecolor=blue0,urlcolor=blue0]{hyperref}
\usepackage{subfigure}

\begin{document}

\title{Maximum mass cutoff in the neutron star mass distribution and the prospect of forming supramassive objects in the double neutron star mergers}

\author{Dong-Sheng Shao}
\affiliation{Key Laboratory of Dark Matter and Space Astronomy, Purple Mountain Observatory, Chinese Academy of Sciences, Nanjing 210033, China}
\affiliation{School of Astronomy and Space Science, University of Science and Technology of China, Hefei, Anhui 230026, China}
\author{Shao-Peng Tang}
\affiliation{Key Laboratory of Dark Matter and Space Astronomy, Purple Mountain Observatory, Chinese Academy of Sciences, Nanjing 210033, China}
\affiliation{School of Astronomy and Space Science, University of Science and Technology of China, Hefei, Anhui 230026, China}
\author{Jin-Liang Jiang}
\affiliation{Key Laboratory of Dark Matter and Space Astronomy, Purple Mountain Observatory, Chinese Academy of Sciences, Nanjing 210033, China}
\affiliation{School of Astronomy and Space Science, University of Science and Technology of China, Hefei, Anhui 230026, China}
\author{Yi-Zhong Fan}
\email[Corresponding author.~]{yzfan@pmo.ac.cn}
\affiliation{Key Laboratory of Dark Matter and Space Astronomy, Purple Mountain Observatory, Chinese Academy of Sciences, Nanjing 210033, China}
\affiliation{School of Astronomy and Space Science, University of Science and Technology of China, Hefei, Anhui 230026, China}

\begin{abstract}
The sample of neutron stars with a measured mass is growing quickly. With the latest sample, we adopt both a flexible Gaussian mixture model and a Gaussian plus Cauchy-Lorentz component model to infer the mass distribution of neutron stars and use the Bayesian model selection to explore evidence for multimodality and a sharp cutoff in the mass distribution. The two models yield rather similar results. Consistent with previous studies, we find evidence for a bimodal distribution together with a cutoff at a mass of $M_{\rm max}=2.26_{-0.05}^{+0.12}M_\odot$ (68\% credible interval; for the Gaussian mixture model). If such a cutoff is interpreted as the maximum gravitational mass of nonrotating cold neutron stars, the prospect of forming supramassive remnants is found to be quite promising for the double neutron star mergers with a total gravitational mass less than or equal to 2.7$M_\odot$ unless the thermal pions could substantially soften the equation of state for the very hot neutron star matter. These supramassive remnants have a typical kinetic rotational energy of approximately $1-2\times 10^{53}$ ergs. Together with a high neutron star merger rate approximately $10^{3}~{\rm Gpc^{-3}~yr^{-3}}$, the neutron star mergers are expected to be significant sources of EeV($10^{18}$eV) cosmic-ray protons.
\end{abstract}

\keywords{Neutron stars$-$Compact binary stars$-$Gravitational waves}

\maketitle

\section{introduction}
The mass distribution of neutron stars (NSs) is very helpful in revealing the mechanism of supernova explosion, the accretion dynamics of binary neutron star, the equation of state of matter with ultrahigh density, the mechanism of cosmic-ray acceleration, etc. Since the discovery by \citet{Hewish1968}, more than 2500 NSs have been detected. The mass measurements, however, are much more challenging and such a goal has just been achieved for a small fraction of these extreme objects, usually in NS binary systems. Nevertheless, benefiting from improvements on pulsar radio timing and x-ray observation, the growing population of the NSs with reliable mass measurements within decades makes it feasible to statistically infer the features of the distribution \citep{Zhang2011, Valentim2011, Ozel2012, Kiziltan2013, Antoniadis2016}. A bimodal distribution with two peaks at approximately $1.3 M_\odot$ and approximately $1.5-1.7 M_\odot$ was suggested in the above literature, which can be well explained by different formation channels and evolution scenarios \citep{Horvath2017}. The other recent intriguing/remarkable finding is a significant cutoff at the high end of the mass distribution ($M_{\rm max}=2.12^{+0.09}_{-0.12}M_\odot$), which is most likely the maximum gravitational mass $(M_{\rm TOV})$ of the nonrotating neutron star\footnote{Though some objects are millisecond pulsars, their rotations are not quick enough to enhance the maximum gravitational mass effectively (see, e.g., Refs. \citep{Friedman1986, Fan2013a, Breu2016}).} and plays an important role in bounding the equation of state (EoS) of NS matter \citep{Alsing2018}. It also directly sets a robust lower limit on the mass of stellar-origin black holes.

Since the publication of \citet{Alsing2018}, rapid progress has been made on the mass measurements of NSs. In particular, several NSs are found to be very massive. For instance, PSR J1600$-$3053 has a mass of $2.3^{+0.7}_{-0.6}M_\odot$ \citep{Arzoumanian2018}, PSR J0740+6620 has a mass of $2.14^{+0.10}_{-0.09}M_\odot$ \citep{Cromartie2020}, PSR J1959+2048 has a mass of $2.18\pm0.09M_\odot$, PSR J2215+5135 has a mass of $2.28^{+0.10}_{-0.09}M_\odot$ \citep{Kandel2020} (please note that the other group reported a mass of $2.27^{+0.17}_{-0.15}M_\odot$ \citep{Linares2018}), and J1811$-$2405 has a mass of $2.0^{+0.8}_{-0.5}M_\odot$ \citep{Ng2020}. In this work the uncertainties are for 68.3\% confidence level unless specifically noticed. Therefore, it is necessary to update the analysis of \citet{Alsing2018} with the latest sample of NSs with mass measurements/information and new fit functions. That is the main motivation of this study.

Our work is structured as follows. In Sec.~\ref{sec:ns-mass-dis} we collect the mass measurements of NSs from the latest literature and then analyze the maximum mass cutoff in the mass distribution. In Sec.~\ref{sec:cutoff-prospect}, as a direct application of the inferred maximum mass cutoff, we adopt an EoS-insensitive approach to examine the prospect of forming supramassive neutron stars (SMNSs) in the double neutron stars(DNS) mergers and discuss the possibility that the EoS softening effect of thermal pions generated in the very hot neutron star can be probed. Motivated by a promising formation prospect, we estimate the kinetic rotational energy of these SMNSs (again, in an EoS-insensitive way) and then their role in accelerating EeV cosmic rays. Finally we summarize our results with some discussions.

\section{updated estimate of the maximum mass cutoff in the neutron star mass distribution}\label{sec:ns-mass-dis}
\subsection{Neutron star mass measurements}
Up to now, almost all the reliable mass measurements were carried out for neutron stars in binary systems. Benefiting from Kepler's Third Law, the orbital parameters of those neutron stars and their companions, which can be measured by either radio timing of pulsations or x-ray/optical observations, make it possible to determine the masses of NSs (see Refs. \citep{Remillard2006, Ozel2016} for recent reviews).

Generally in the Newtonian frame, we can measure five Keplerian parameters for the orbital motions of binary: the binary period $P_{\rm b}$, the eccentricity $e$, the component of the pulsar's semimajor axis $a_{\rm p}$ along the line of sight $x_{\rm p} = a_{\rm p} \sin{i}/c$ (where $i$ is the orbital inclination angle and $c$ is the speed of light), and the time and longitude of periastron $T_0$ and $\omega$. Then, the so-called mass function of the binary, defined as $f\equiv \frac{(m_{\rm c} \sin{i})^3}{(m_{\rm p}+m_{\rm c})^2}=\left(\frac{2\pi}{P_{\rm b}}\right)^2\frac{x_{\rm p}^3}{G}$ \citep{Remillard2006}, is dependent on $P_{\rm b}$ and $i$, where $m_{\rm p}$ and $m_{\rm c}$ stand for the masses of pulsar and its companion respectively, and $G$ is the Newtonian gravitational constant. The degeneracies of the unknowns can be broken as long as the mass ratio $q\equiv m_{\rm p}/m_{\rm c}=x_{\rm c}/x_{\rm p}$ (where $x_{\rm c}$ is the projection of the companion's semimajor axis on the line of sight) and the mass of the companion are determined, as briefly summarized below (see Ref.\citep{Ozel2016} and the references therein). For some binary systems with a main-sequence star or bright white dwarf companion, $q$ and $m_{\rm c}$ can be measured by studying the spectrum of the companion (e.g., the Balmer line of hydrogen in the atmosphere via phase-resolved optical spectroscopy), leading to a reliable mass measurement. For double NS systems (or binaries with massive white dwarf companion), the components are compact enough to make the relativistic effects on the orbital motion observation. These effects, described by five post-Keplerian (PK) parameters, including the periastron precession $\dot\omega$, Einstein delay $\gamma$, the shape and range of the Shapiro delay $s$ and $r$, and the orbital period decay $\dot P_{\rm b}$, depend sensitively on the masses of components and the Keplerian parameters of their orbit (see Ref.\citep{Stairs2003} and references therein). Once some of them have been precisely measured by radio timing of pulsars, the individual mass can be determined with the least uncertainty, especially when both of the components happened to be pulsars. For neutron stars with high or low stellar mass companion, the observations from x-ray/optical bands provide a viable approach to determine the orbital parameters and the masses. The measurement of eclipsing of the x ray from the high-mass x-ray binary (HMXB) yields some fundamental parameters of binary orbit, such as the period $P_{\rm b}$, the eccentricity $e$, the longitude of periastron $\omega_0$, and the semimajor axis of the neutron star's orbit $a_{\rm x} \sin{i}$. Together with the information of velocity and inclination of companion obtained from optical observations, the mass can be solved from the basic equations. Modeling the thermal emission of neutron star atmosphere can constrain both mass and radius of quiescent low-mass x-ray binary (qLMXB). The thermonuclear x-ray burst, a helium flash occurring in the surface layer of the accreting neutron star of the low-mass x-ray binary(LMXB), can also be used to measure mass and radius, by combining analysis of the apparent angular size, the Eddington flux, and the source distance. Most events included in our sample (Table\ref{tab:mass_data}) are measured in the above ways.

For the merging NSs in the deep universe, their masses can be measured with the gravitational wave data. Until January 2020, only two neutron star merger events had been reported. The GW170817 data strongly favored the double NS merger origin \citep{GW170817}, while for GW190425, a NS-black hole binary system \citep{Han2020, Foley2020} cannot be ruled out. In this work following the LIGO and Virgo collaborations  \citep{GW190425}, we attribute it to the merger of a pair of relatively massive NSs. The double neutron star merger sample, though still small right now, is expected to increase rapidly in the next decade. For completeness we include these objects in our analysis. As for the isolated neutron stars, the pulse-profile modeling can simultaneously yield the masses and radii of some nearby bright millisecond pulsars. With the NICER data, such a goal was achieved first for PSR J0030+0451 \citep{Riley2019, Miller2019}. This source is also included in Table\ref{tab:mass_data}. Very recently, \citet{Tang2020} proposed a new method to infer the masses of a few isolated neutron stars with the gravitational redshift measurements. Though interesting, such an approach is model dependent, and we do not include these events in our sample.

We updated the sample listed in Table 1 of \citet{Alsing2018} in two aspects. First, the events with improved mass measurements have been updated. Second, the new events with mass measurements, dated from April 2018, including a few very massive ones such as PSR J0740+6620, \citep{Cromartie2020}, J1959+2048 and J2215+5135 \citep{Kandel2020}, and J1811$-$2405 \citep{Ng2020}, have been added. Five NSs with the masses measured by LIGO and NICER missions have also been included. In comparison to \citet{Alsing2018}, the total number of NSs in our sample increased from 74 to 103 (see Table\ref{tab:mass_data} in the Appendix for details).

\subsection{Evidence for a maximum mass cutoff in the neutron star mass distribution: Updated analysis}

\begin{figure}[!ht]
\begin{center}
\includegraphics[width=0.9\textwidth]{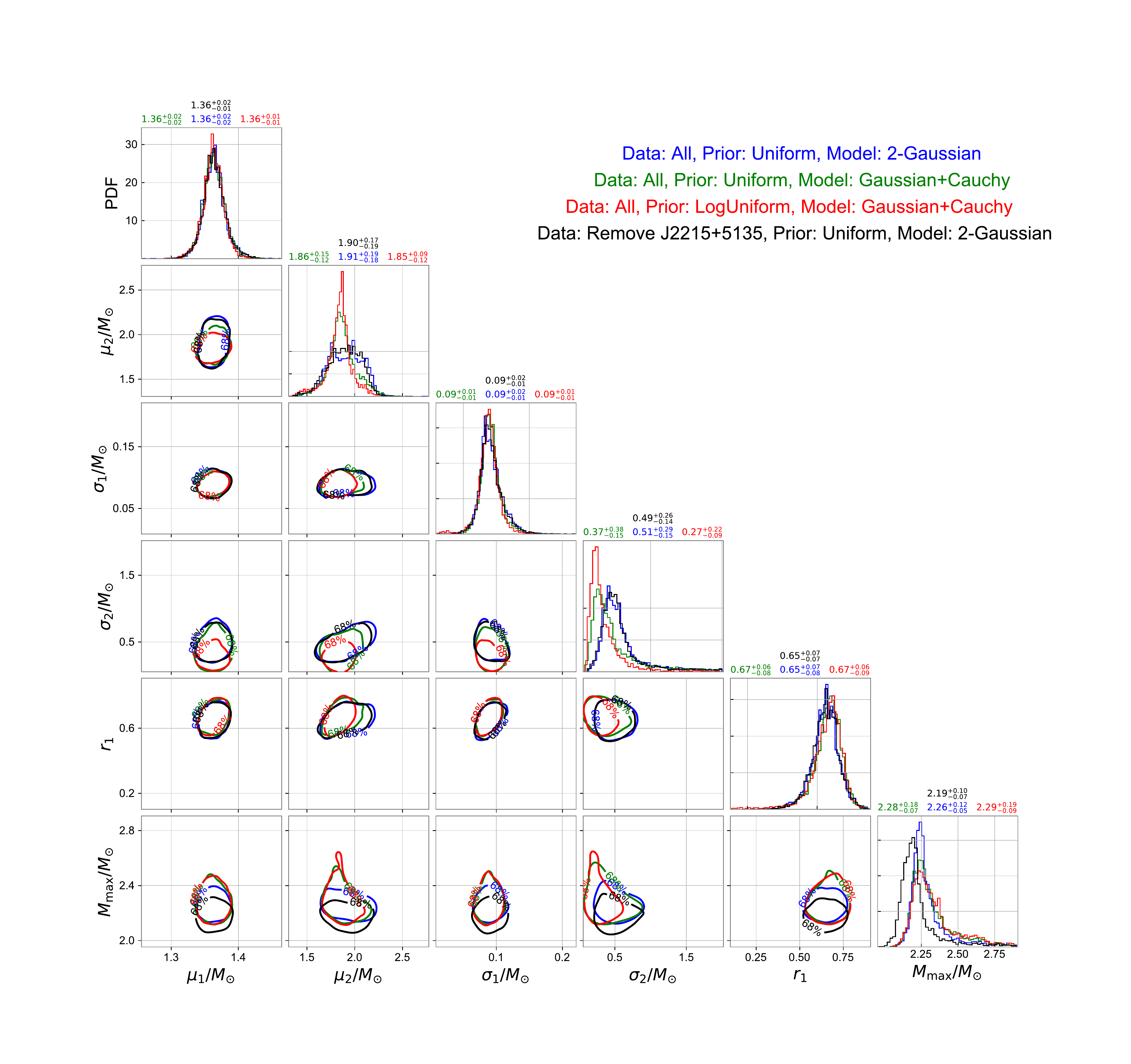}
\end{center}
\caption{Distributions of parameters ($\vec{\theta}$) of the NS mass distribution obtained with some different priors and models.}
\label{fig:mmax}
\end{figure}{}

Benefiting from the accumulated NS mass measurements mentioned above, we can statistically investigate the properties of NS mass distribution. Previous works have proposed various models of NS population to fit the observation data, such as uniform, single Gaussian, bimodal Gaussian \citep{Farrow2019}, multicomponent Gaussian \citep{Alsing2018}, and skewed normal distribution \citep{Kiziltan2013}. In this work, we use two models to fit the latest sample. The first is a two-component Gaussian mixture model with a maximum mass cutoff $M_{\rm max}$, i.e.,
\begin{equation}
\label{eq:mixture}
P(m_\mathrm{p}\mid \vec{\theta}) = r_1 \mathcal{N}_1(m_\mathrm{p}\mid \mu_1, \sigma_1)/\Phi_1 + (1-r_1) \mathcal{N}_2 (m_\mathrm{p}\mid \mu_2, \sigma_2)/\Phi_2, ~ {\rm for} ~ m_\mathrm{p} \in [M_{\rm min}, M_{\rm max}],
\end{equation}
where $\vec{\theta} = \{\mu_1, \mu_2, \sigma_1, \sigma_2, r_1, M_{\rm max}\}$, and inside the brackets, $\mu_1(\mu_2)$, $\sigma_1(\sigma_2)$, $r_1$, and $m_\mathrm{p}$, respectively, denote the mean, standard deviation, relative weight of the two components, and the pulsar mass. The second model is a mixture of Gaussian component $\mathcal{N}_1$ and a Cauchy-Lorentz component ${\rm Ca}_2$, whose function form is given by replacing the second Gaussian component $\mathcal{N}_2$ of Eq.(\ref{eq:mixture}) with ${\rm Ca}_2$; thus, $\mu_2$ and $\sigma_2$ represent the location and scale parameters of Cauchy-Lorentz distribution. In our analysis, we set $M_\mathrm{min}=0.9M_\odot$, which is a reasonable lower bound to contain most of the NS mass measurements to probe the distribution properties. The normalization constants $\Phi_i~(i=1,2)$ are integrals over each component $(\cdot)_i$ (over the allowed NS mass range) using $\Phi_i = \int_{M_\mathrm{min}}^{M_\mathrm{max}} (\cdot)_i(x\mid \mu_i, \sigma_i) \mathrm{d} x$. To approximate well the non-Gaussian mass measurements in Table\ref{tab:mass_data}, we use the asymmetric normal distribution studied in Refs.\citep{Kiziltan2013,Fernandez:98} to reproduce the error distribution, of which the density function is given by
\begin{eqnarray}\label{skew-normal-errors}
{\rm AN}(w\mid c,d)=\frac{2}{d(c+\frac{1}{c})} \left\{ \phi\left(\frac{w}{cd}\right)1_{[0,\infty)}(w) + \phi\left(\frac{c w}{d}\right)1_{(-\infty,0)}(w) \right\},
\end{eqnarray}
where $c > 0$, $d > 0$, $\phi$ means normal distribution, and $1_{A}(\cdot)$ denotes the indicator function of set $A$. Thus given the NS mass measurements ${\mathcal{M}_{i}}_{-\ell_{i}}^{+u_{i}}$ ($+u_{i}/-\ell_{i}$ are 68\% central limits), parameters $c_{i}$ and $d_{i}$ for the $i$th NS can be estimated through $c_{i}=$ $(u_{i}/\ell_{i})^{1/2}$ and $\int_{-\ell_{i}}^{u_{i}} \text{AN}(w\mid c_{i},d_{i}) \text{d}w = 0.68$. Then it is straightforward to calculate the probability for a specific pulsar mass $m_{\rm p}$ via $P(D^i\mid m_{\rm p})=\text{AN}(m_{\rm p}-\mathcal{M}_{i}\mid c_{i}, d_{i})$. For the data only having mass function $f$ and mass ratio $q$ (or total mass $m_{\rm T}$) measurements available in Table\ref{tab:mass_data}, we adopt the Eqs.(3) and (4) of \citet{Alsing2018} to evaluate the probability $P(D^i\mid m_{\rm p})$. Therefore, the likelihood constructed for our inference is given by
\begin{equation}
\label{eq:likelihood}
L(D\mid \vec{\theta}) \propto \prod_{i=1}^N \int P(m_\mathrm{p} \mid \vec{\theta}) P(D^i \mid m_{\rm p}) {\rm d} m_\mathrm{p},
\end{equation}
with which we can use the nest sampling technique, e.g., {\sc PyMultiNest} sampler \citep{2016ascl.soft06005B}, to obtain the samples of the parameters ($\vec{\theta}$) of NS mass distribution. The ranges of $\vec{\theta}$ are chosen as follows: $\mu_i \in [0.9, 2.9]M_\odot$, $\sigma_i \in [0.01, 2]M_\odot$, $r_1 \in [0.1, 0.9]$, and $M_{\rm max} \in [1.9, 2.9]M_\odot$. We take both uniform and uniform-in-log priors to perform nest sampling, and further request that $\mu_1<\mu_2<M_{\rm max}$ and $\sigma_1<\sigma_2$.

\begin{figure}[!ht]
\begin{center}
\includegraphics[width=0.45\textwidth]{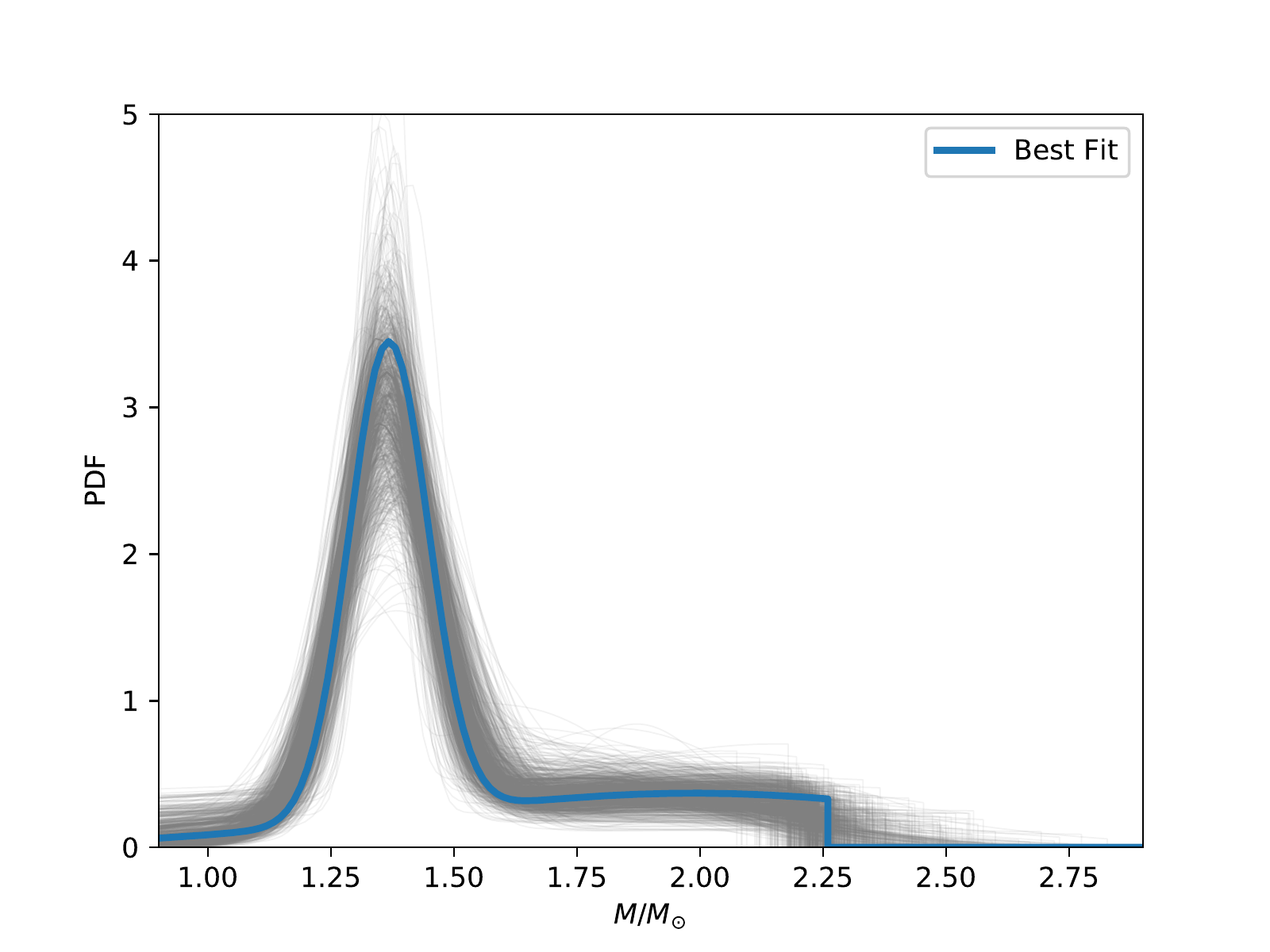}
\includegraphics[width=0.45\textwidth]{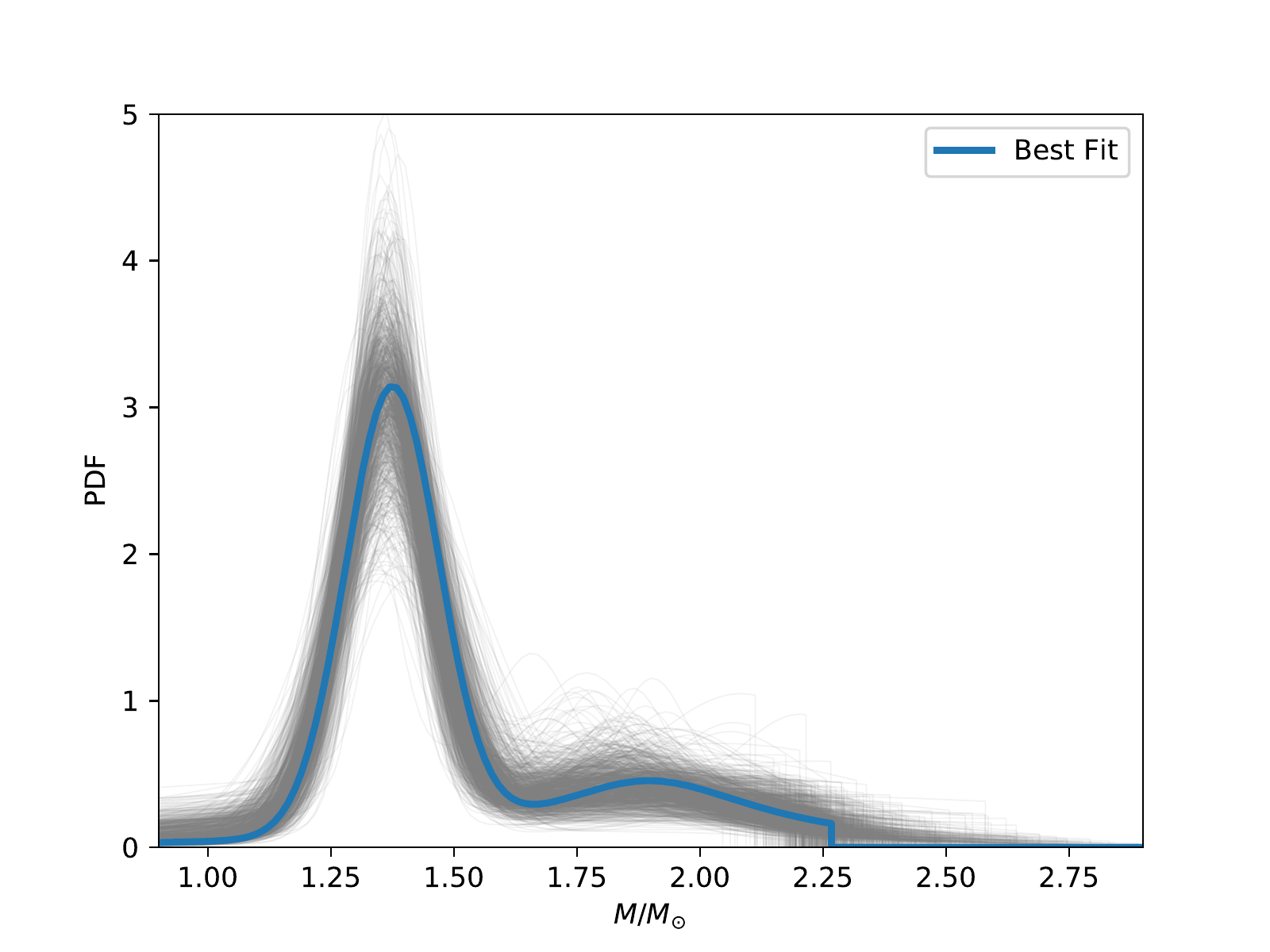}
\end{center}
\caption{Maximum $a posteriori$ NS mass distribution (blue) with 1000 independent posterior samples to give a visual guide for the uncertainties, under the considered model that is most preferred by the data. Left panel: the two-component Gaussian mixture with a sharp cutoff of approximately $2.26M_\odot$ (median value). Right panel: the mixture of Gaussian and Cauchy-Lorentz components with a sharp cutoff of approximately $2.28M_\odot$ (median value).}
\label{fig:mmax-post}
\end{figure}{}
In Bayes's statistic frame, the method to evaluate the model preference is the odds ratio of the probability of two hypotheses which is given by
\begin{equation}
\mathcal{O}_{12} = \frac{P({\rm D}\mid \mathcal{H}_1)}{P({\rm D}\mid \mathcal{H}_2)}\frac{P(\mathcal{H}_1)}{P(\mathcal{H}_2)},
\end{equation}
where $P({\rm D}\mid \mathcal{H})$ is the Bayesian evidence (or marginal likelihood) $\mathcal{Z}$ for a given model hypothesis $\mathcal{H}$, $P({\rm D}\mid \mathcal{H}_1)/P({\rm D}\mid \mathcal{H}_2)$ is the Bayes factor $\mathcal{K}_{12}$, and $P(\mathcal{H}_1)/P(\mathcal{H}_2)$ is the prior odds ratio that defines our prior relative belief in these two models (here we set to unity as an $a priori$ agnostic). To estimate the ``evidence" of the maximum mass cutoff, we follow the procedure of \citet{Alsing2018} by comparing the evidence to the model of fixing $M_{\rm max}=2.9\,M_\odot$ (i.e., without mass cutoff). For the two-component Gaussian model, the logarithm evidence of with (without) cutoff is $-3.47~(-5.38)$, and for the mixture of Gaussian and Cauchy-Lorentz model, it is $-3.04~(-5.02)$, which show a positive support for maximum mass cutoff \citep{Kass1995}, consistent with the previous work \citep{Alsing2018}.

Some results are reported in Fig.\ref{fig:mmax}, and for two-component Gaussian model, we have $M_{\rm max}=2.26_{-0.05}^{+0.12}M_\odot$ (68\% credible interval; the 95\% credible interval is $M_{\rm max}=2.26_{-0.11}^{+0.47}M_\odot$). We have also tested the removal of one or more very massive neutron stars whose mass measurement methods are not so direct/widely accepted, e.g., PSR J2215+5135 and PSR J1959+2048. Without PSR J2215+5135, we have $M_{\rm max}=2.19^{+0.10}_{-0.07}~M_\odot$. If we further remove PSR J1959+2048, the result then becomes $M_{\rm max}=2.13^{+0.12}_{-0.07}~M_\odot$, while for the mixture of Gaussian and Cauchy-Lorentz model, we have $M_{\rm max}=2.28_{-0.07}^{+0.18}M_\odot$, $2.21_{-0.08}^{+0.10}M_\odot$, and $2.15_{-0.08}^{+0.12}M_\odot$ for the full data and the sequential removal of PSR J2215+5135 and PSR J1959+2048, respectively. Our results show that the choices of priors and NS mass distribution models have little influence on bounding $M_{\rm max}$; the inclusion of ``heavier" NSs can effectively shift the bounds on $M_{\rm max}$. In view of these facts and considering that the mass measurements of Ref.\citep{Kandel2020} may suffer from some systematic uncertainties, for a cross-check we instead adopt the mass of $2.27_{-0.15}^{+0.17}M_\odot$ for PSR J2215+5135 measured by \citet{Linares2018} in our modeling and infer the $M_{\rm max}$ to be $2.22_{-0.06}^{+0.10}M_\odot$ (two-Gaussian-component model) and $2.22_{-0.07}^{+0.13}M_\odot$ (Gaussian plus Cauchy-Lorentz component model), respectively. Since they are very consistent with the result using the data of Ref.\citep{Kandel2020}, we still take our results based on the ``latest" mass measurement sample as the fiducial ones. We show in Fig.\ref{fig:mmax-post} the maximum $a posteriori$ mass distribution for the two models with free $M_{\rm max}$ with 1000 independent posterior samples plotted over the top to give a visual impression of the uncertainties on the shape of the distribution. We take the best-fit result of the two-component Gaussian mixture (shown in the blue line) as our fiducial model.

\begin{figure}[!ht]
\begin{center}
\includegraphics[width=0.4\textwidth]{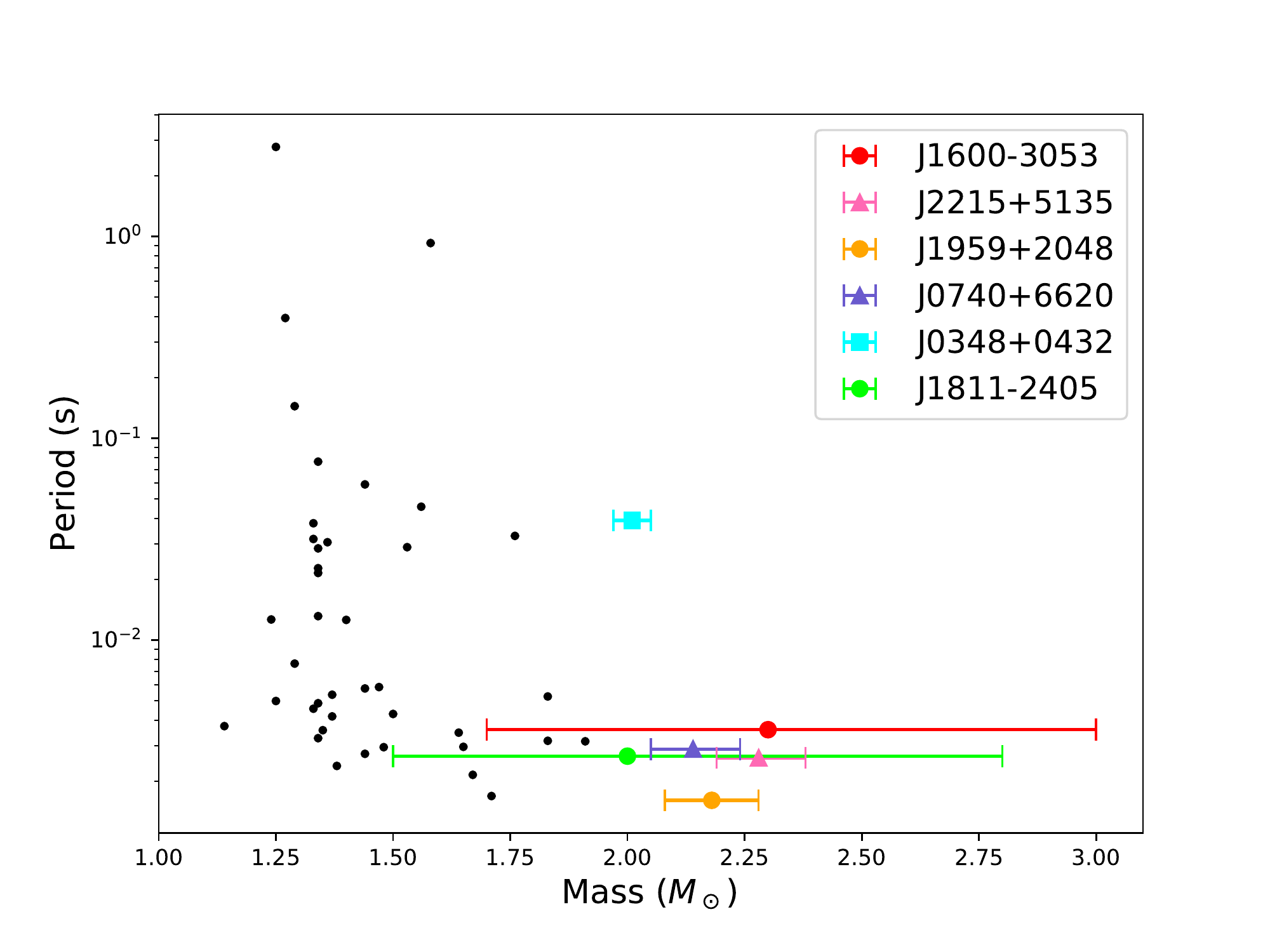}
\includegraphics[width=0.4\textwidth]{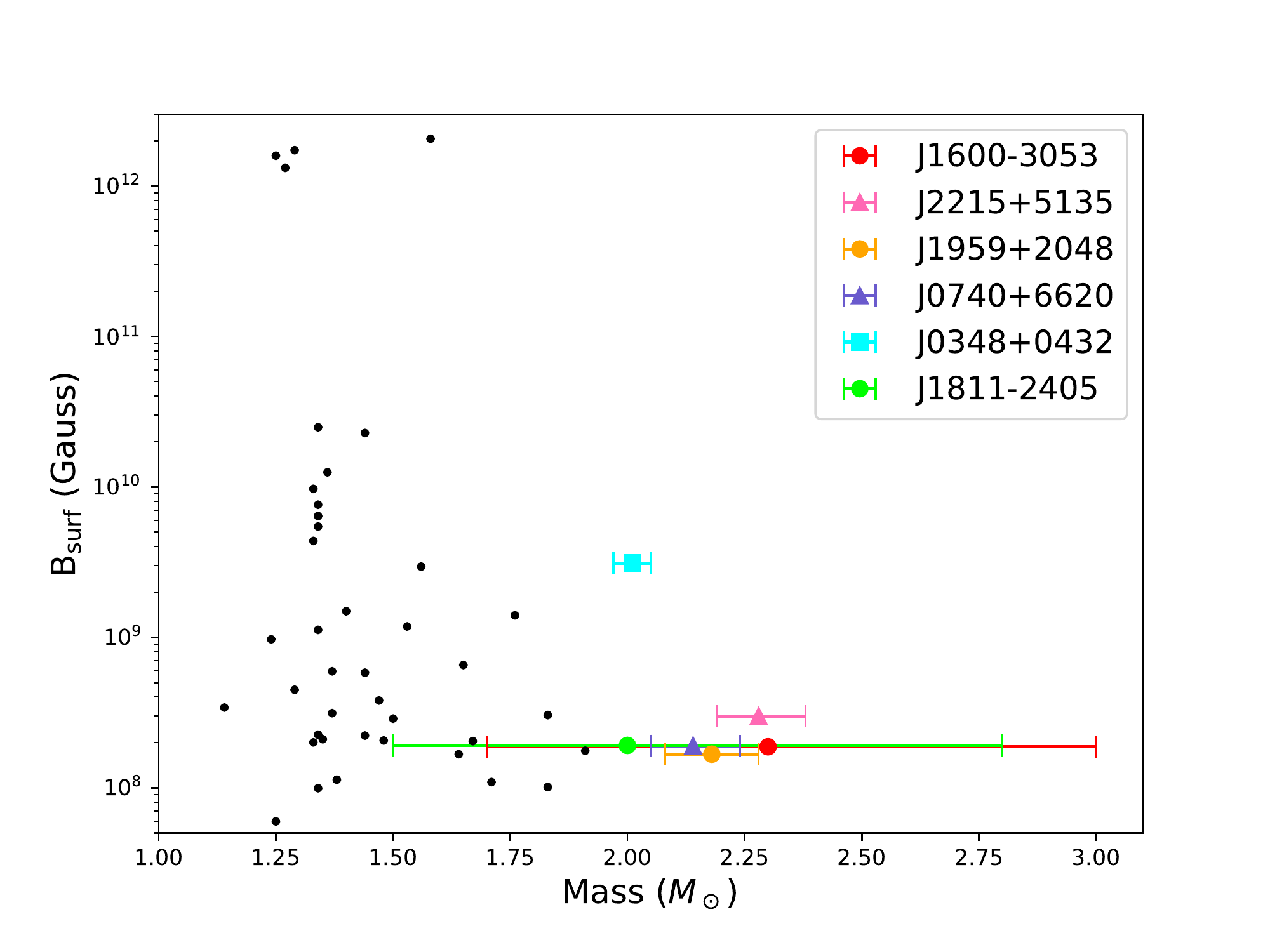}
\includegraphics[width=0.4\textwidth]{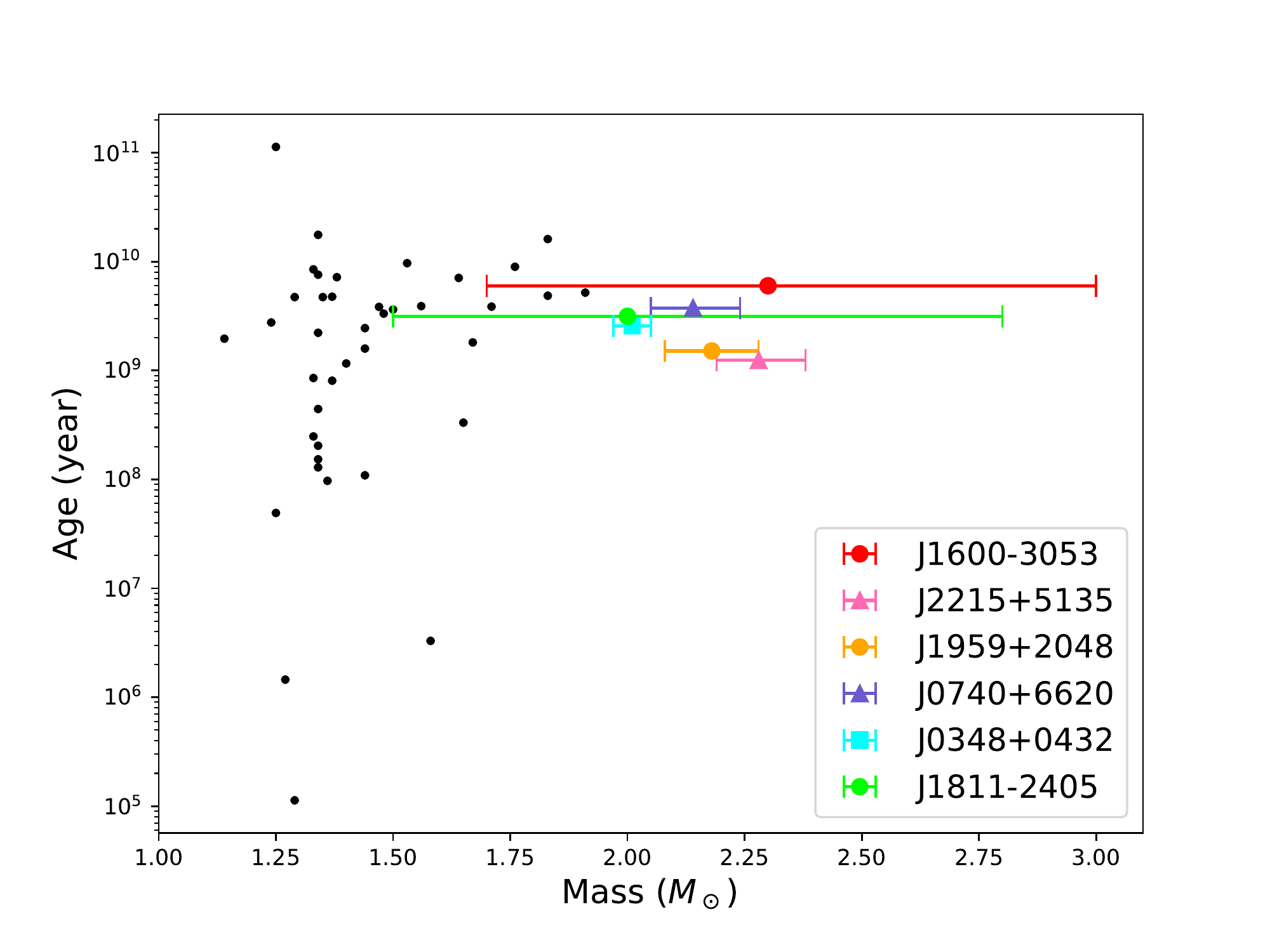}
\includegraphics[width=0.4\textwidth]{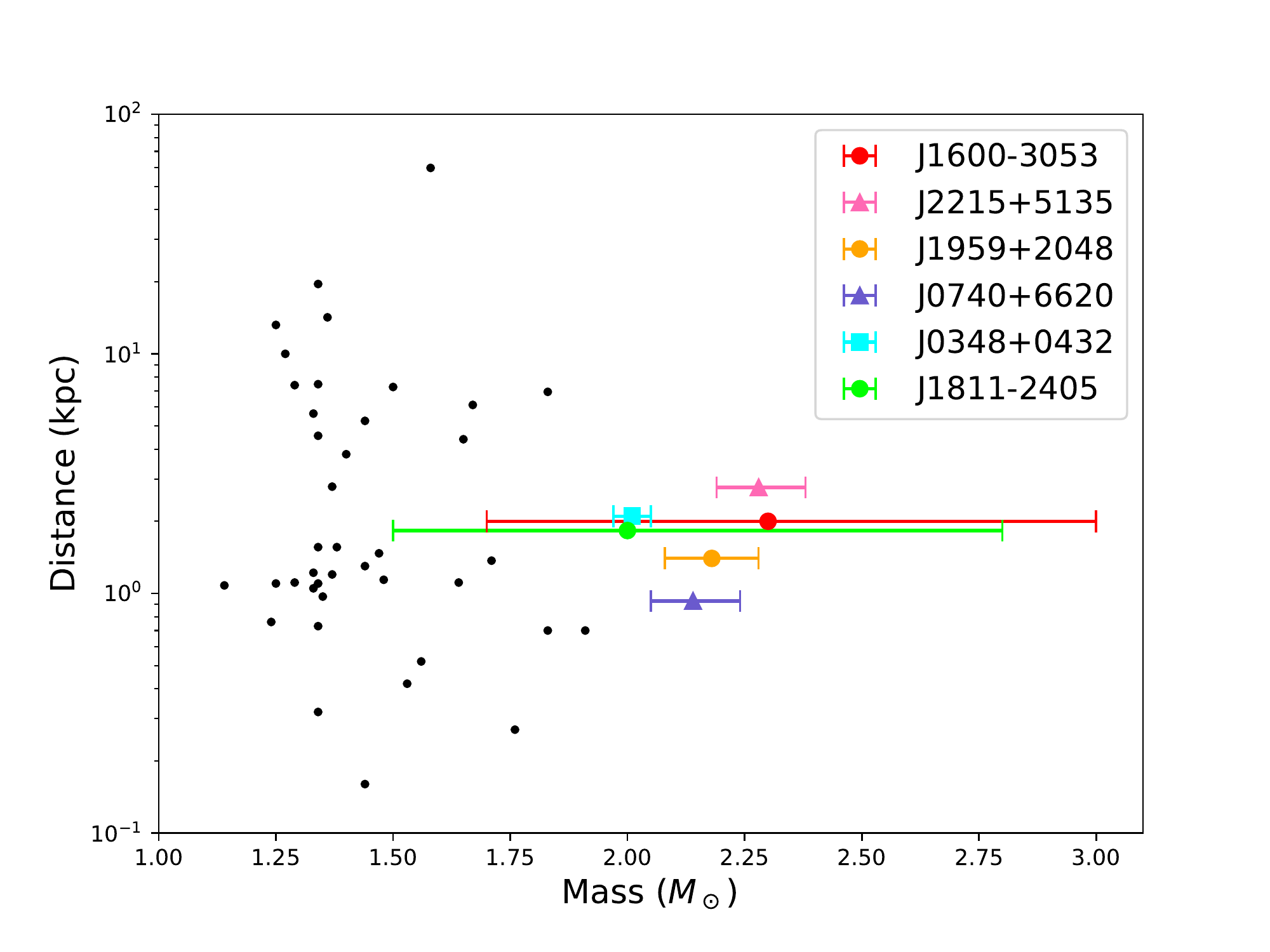}
\end{center}
\caption{Statistical characteristics of pulsar mass vs spin period, magnetic field strength on surface, age and distance. The most massive neutron stars (which may be heavier than 2.0$M_\odot$) are displayed in colored circles, triangles, and squares with 1$\sigma$ error bar.}
\label{fig:mass_pulsar}
\end{figure}

\begin{figure}[!ht]
\begin{center}
\includegraphics[width=0.7\textwidth]{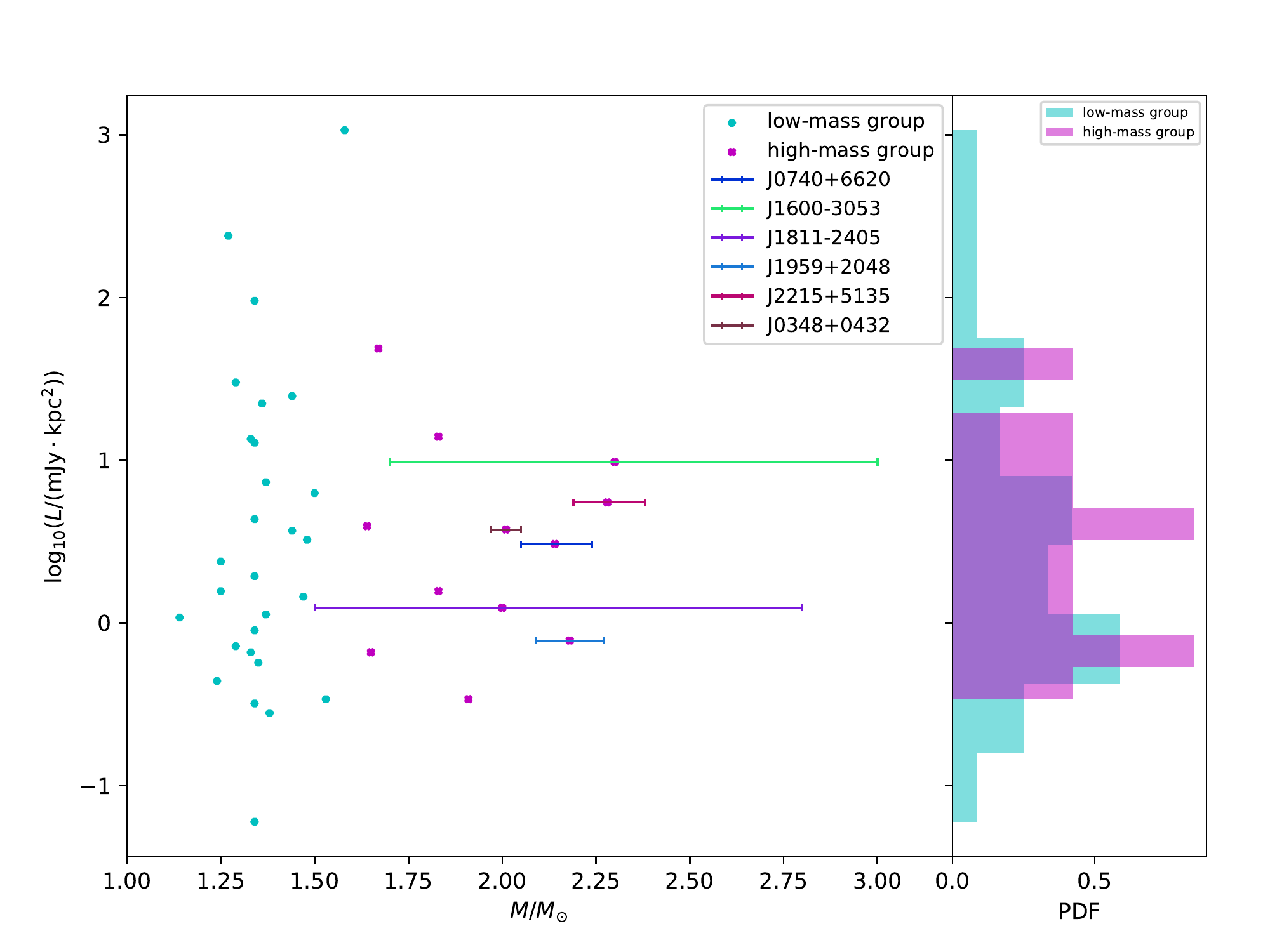}
\end{center}
\caption{Pulsar mass vs radio luminosity. The massive neutron stars  (which may be heavier than 1.6$M_\odot$) are displayed in magenta crosses with 1$\sigma$ error bar. The radio luminosities were measured at 1.4 (most events) or 0.35GHz (a few objects) \citep{ATNF} except J0348+0432 at 0.82GHz \citep{Lynch2013} and J1807-2500B at 2~GHz \citep{Lynch2012}. All the luminosities have been normalized to 1.4GHz (if the spectral index is unavailable we take the mean value of $-1.4$ \citep{Bates2013}).}
\label{fig:mass_lumi}
\end{figure}

Note that our sample, though significantly extended in comparison to \citet{Alsing2018}, still suffers from some selection effects. In particular, almost all the mass data available come from NSs in binaries, and all the most precise mass measurements come from double NS systems; there is still no strong evidence yet that these events represent well the whole population of neutron stars. Anyhow, a very recent study shows that the neutron stars, even born in the death of the very massive stars with a zero metallicity, have a gravitational mass below $1.8M_\odot$ \citep{Ebinger2020}. Therefore, it is unlikely to find some isolated NSs as massive as $M_{\rm max}\sim 2.2M_\odot$. (Indeed, the current measurement/estimates of the isolated NSs find a low mass of approximately $1.2-1.5M_\odot$ \citep{Riley2019,Tang2020}, though the sample is still quite small.) Moreover, we have shown that the heaviest objects rather than the precisely measured neutron star masses in the double neutron star binaries play the key role in governing $M_{\rm max}$. The more relevant selection effects may arise from the observations/identifications of the neutron star binary systems.
To examine this possibility we have collected some measured/derived properties of our pulsar sample from the ATNF (Australia Telescope National Facility) catalog \citep{ATNF}.\footnote{http://www.atnf.csiro.au/research/pulsar/psrcat/, version 1.63.} These properties include the spin period, magnetic field strength on the surface, age, and distance available for 46 objects with mass measurements. For the radio luminosity and flux, the sample is a bit smaller. Nevertheless, these most massive events (except Vela X1 which was detected in optical and x-rays) have been included. Below, we focus on the objects that are possibly more massive than $2M_\odot$ (see the colored points in Fig.\ref{fig:mass_pulsar}).  All these very massive objects are rather old (greater than or equal to $10^{9}$ yrs) and have very low surface magnetic fields (approximately a few $10^{8}~{\rm G}$; the ``exception" is PSR J0348+0432 which has a ``relatively" high magnetic field strength of $3.1\times 10^{9}~{\rm G}$). Their rotation periods are short (less than $10~{\rm ms}$, except PSR J0348+0432 with a rotation period of $39.1~{\rm ms}$ that may be caused by its relatively high surface magnetic field). These properties are very consistent with the recycle nature of such objects.  The recycle (i.e., significant matter-accretion) experience of these objects is also supported by the type of their companion stars (see Table 1 for a complete list).
Intriguingly, both PSR J0348+0432 (the most accurately measured object in the very massive NS sample besides PSR J0348+0432) and PSR J2215+5135 (possibly the most massive one detected so far) are among the luminous radio pulsars (see Fig.\ref{fig:mass_lumi}). Therefore, at least for the current sample, we do not find evidence/indication for a sizeable nondetecion/misidentification probability of the very massive NSs in radio. Considering the above facts, in agreement with \citet{Alsing2018}, we suggest that some selection effects might leave some imprint on the inferred mass distribution, while it seems unlikely that they are responsible for the inferred hard cutoff at $M_{\rm max}$.
In order to further check whether the radio emission plays an important role in shaping the mass distribution, we have further carried out a Chi-square test of the independence between the masses and radio luminosities of the current sample and got a p-value as low as $6.3\times 10^{-9}$, suggesting no linear correlation among the masses and luminosities.  To check potential nonlinear correlation between the two variables, we have also fitted the data with a multivariate adaptive regression code called {\sc pyearth}\citep{Friedman1991},\footnote{https://contrib.scikit-learn.org/py-earth/content.html.}which is effective at identifying a linear or nonlinear relation. The best fit is a constant function, indicating the absence of any linear or nonlinear correlation between the mass and luminosity. This is also evident in Fig.\ref{fig:mass_lumi} which displays the masses and radio luminosities of $40$ pulsars. In the right part we show the distribution of luminosities of two groups of NSs separated by $M=1.6M_{\odot}$, namely the high-mass group and the low-mass group. The two sided Kolmogorov-Smirnov test has been adopted to examine whether the two groups of luminosities share the same intrinsic distribution or not and we have got a p-values of $0.96$, which favors the same intrinsic luminosity distribution hypothesis (a highly relevant/consistent conclusion was also drawn in Ref.\citep{Szary2014}).
We there ore conclude that there is no evidence for a correlation between the mass and radio luminosity of the neutron stars. As a result of the lack of identification of the mass-dependent selection effects (see, e.g., Refs. \citep{Lorimer2006,Kiziltan2013} for previous investigations), we do not expect that these selection effects will introduce serious bias to the observed mass distribution (see also Ref.\citep{Kiziltan2013}).

{}

{}

\section{Supramassive neutron stars: formation prospect in double NS mergers and the acceleration of EeV cosmic rays}\label{sec:cutoff-prospect}
{The inferred $M_{\rm max}$ can be used to tighten the bounds on the equation of state of the extreme dense matter of the neutron stars (\citep[e.g., Ref.][]{Alsing2018}).
The other direct/important application is to estimate the fate of the remnants
formed in the double neutron star mergers. As shown below, we find a promising prospect of forming supramassive neutron stars (SMNSs) which has a typical
kinetic rotational energy approximately $1-2\times 10^{53}$ erg. With a reasonably high surface magnetic field (i.e., greater than or equal to $10^{12}$ G), the release of such a huge amount of energy into the surrounding material is quick and the driven energetic blast wave can accelerate the EeV cosmic rays effectively.}

\subsection{Prospect of forming supramassive neutron star in double neutron star mergers}{\label{Sec:SMNS}}
The fates of the remnants formed in the double NS mergers have been extensively examined before. Since the EoS of NS matter is essentially unknown, the relevant examinations were based on some representative models (e.g., Refs.\citep{Morrison2004, Hotokezaka2013a, Lawrence2015, Piro2017, Ma2018}). Now we reexamine the general prospect of forming SMNSs in the double NS mergers with some EoS-insensitive relationships.

The remnants formed in the mergers of the heaviest double NS binary systems may collapse promptly into black holes. But in most cases the initial remnant should be a differentially rotating very massive NS. The differential rotation may be terminated quickly (in a timescale of approximately $0.1-1$ s, as suggested, for example, in Ref.\cite{Hotokezaka2013a}) and the SMNS or even stable NS may be formed (if the uniform rotation cannot support the massive object, it will collapse. We call such a kind of transient objects as the hypermassive NS). Benefiting from some updated EoS-independent relations of NSs, \citet{Shao2020} derived an empirical relation among the critical total gravitational mass of DNSs ($M_{\rm tot,c}$), the mass and compactness of NS in the nonrotation maximum equilibrium configuration (denoted by $M_{\rm TOV}$ and $\zeta_{\rm TOV}$, respectively), the dimensionless angular momentum of remnant at the onset of collapse ($j_{\rm c}$), and the total mass lost apart from the remnant core ($m_{\rm loss}$, including the kilonova/macronova ejecta and the accretion torus/disk), which reads
\begin{equation}
M_{\rm tot,c}\approx M_{\rm TOV}(1+0.079\zeta_{\rm TOV}^{-1}j_{\rm c}^2+0.017\zeta_{\rm TOV}^{-2}j_{\rm c}^4)(0.798+0.971\zeta_{\rm TOV})(1-0.091~M_\odot^{-1}~m_{\rm loss})+m_{\rm loss},
\label{eq:M_tot_c}
\end{equation}
where $j_{\rm kep}$ is the dimensionless angular momentum of NS rotating at Keplerian velocity.
\begin{figure}[!ht]
\begin{center}
\includegraphics[width=0.7\textwidth]{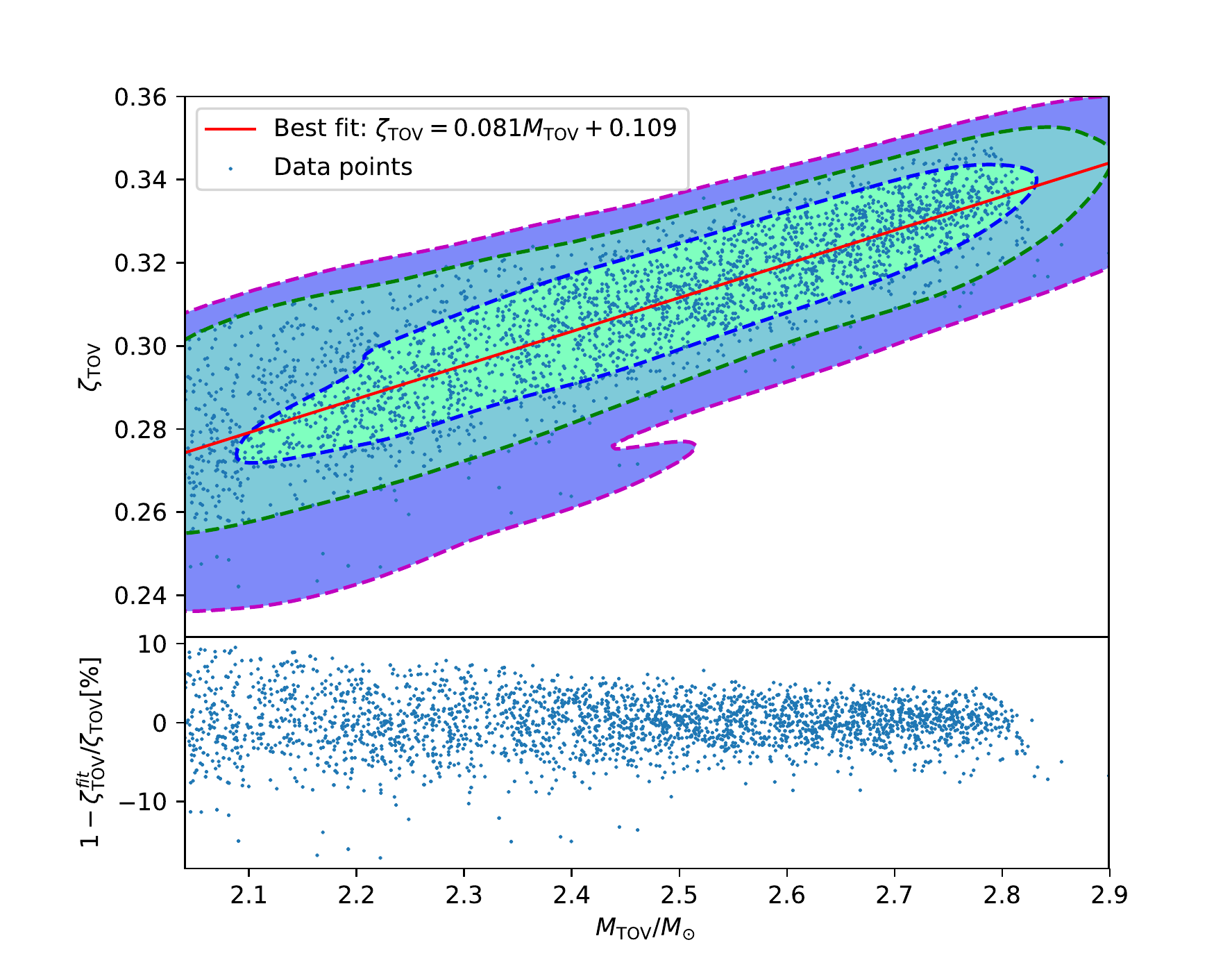}
\end{center}
\caption{$M_{\rm TOV}-\zeta_{\rm TOV}$ correlation inferred from the joint constraints set by the data of PSR J0030+0451, GW170817, and some nuclear data (see Ref.\citep{Jiang2020} for the technical details). The red solid line shows the best-fit model. The blue dashed line, the green dashed line, and the magenta dashed line represent the $1\sigma$, $2\sigma$, and $3\sigma$ contours, respectively.}
\label{fig:MR_tov}
\end{figure}

\begin{figure}[!ht]
\begin{center}
\includegraphics[width=0.7\textwidth]{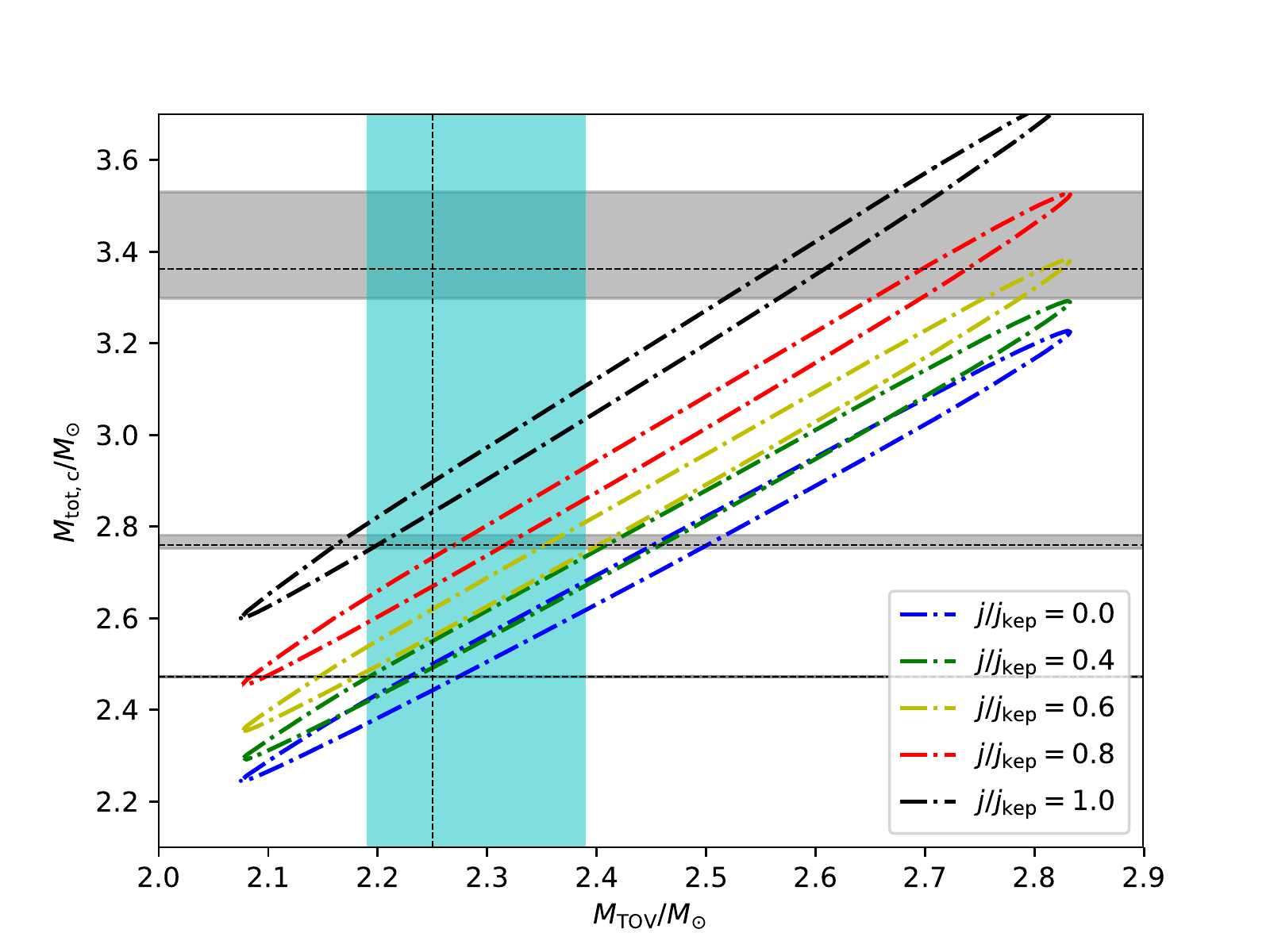}
\end{center}
\caption{Critical total gravitational mass of DNSs vs $M_{\rm TOV}$, supposing $M_{\rm loss} = 0.05 M_{\odot}$. The vertical cyan region represents our evaluated mass cutoff in the NS mass distribution (see Fig.\ref{fig:mmax}). The horizontal gray regions, from the top to the bottom, represent the total gravitational mass of GW190425, GW170817, and PSR J0514$-$4002A (so far, the lightest double NS system identified in the Galaxy), respectively. Note that the regions shown in this plot are for the $68\%$ confidence intervals.}
\label{fig:Mtotc}
\end{figure}
With the empirical relation of $j_{\rm kep}\approx 1.24 \zeta_{\rm TOV}^{0.5}$, Eq.(\ref{eq:M_tot_c}) reduces to
\begin{equation}
M_{\rm tot,c}\approx M_{\rm TOV}\left[1+0.122\left(\frac{j_{\rm c}}{j_{\rm kep}}\right)^2+0.040\left(\frac{j_{\rm c}}{j_{\rm kep}}\right)^4\right](0.798+0.971\zeta_{\rm TOV})(1-0.091~M_\odot^{-1}~m_{\rm loss})+m_{\rm loss}.
\label{eq:M_tot_c_new}
\end{equation}
The merger remnant, if not supported by the thermal pressure and if it has entered the phase of uniform rotation, will collapse to a black hole as
\begin{equation}
M_{\rm tot}>M_{\rm TOV}(0.927+1.129\zeta_{\rm TOV})(1-0.091~M_\odot^{-1}~m_{\rm loss})+m_{\rm loss},
\end{equation}
where $M_{\rm tot}$ is the total gravitational mass of the double NS system.
A stable massive NS will be the output if instead we have
\begin{equation}
M_{\rm tot}<M_{\rm TOV}(0.798+0.971\zeta_{\rm TOV})(1-0.091~M_\odot^{-1}~m_{\rm loss})+m_{\rm loss}.
\end{equation}
For the rest, a SMNS remains as long as the decreasing angular momentum meets the condition of $j\geq j_{\rm c}$, where
\begin{equation}
j_{\rm c}=\sqrt{-1.5+\sqrt{2.3+25.0[{\cal C}^{-1}(M_{\rm tot}-m_{\rm loss})-1.0]}}j_{\rm kep}
\label{eq:jc}
\end{equation}
and ${\cal C}=M_{\rm TOV}(0.798+0.971\zeta_{\rm TOV})(1-0.091~M_\odot^{-1}~m_{\rm loss})$.

In addition to $j_{\rm c}/j_{\rm kep}$ and $m_{\rm loss}$, $M_{\rm TOV}$ and $R_{\rm TOV}$ (note that $\zeta_{\rm TOV}\equiv GM_{\rm TOV}/R_{\rm TOV}c^{2}$) play some roles in shaping $M_{\rm tot,c}$. These two parameters are still unknown yet. Anyhow reasonable constraints have been set by the jointed data analysis of gravitational wave, NS observations, and some nuclear experiments. We take the very recent constraints obtained in the analysis of PSR J0030+0451, GW170817, and some nuclear data \citep{Jiang2020}. The results of the piecewise parametrization approach are adopted here (the spectral parametrization approach yields similar result) and the $M_{\rm TOV}$ is restricted in the range of $(2.04,~2.9)M_\odot$. The two-dimensional probability distribution is shown in Fig.\ref{fig:MR_tov}, with which it is straightforward to predict $M_{\rm tot,c}$ (more exactly, the range) for the given $M_{\rm TOV}$, $m_{\rm loss}$, and $j/j_{\rm kep}$. The results are shown in Fig.\ref{fig:Mtotc}. Clearly, for some light double neutron star binary systems (note that about half of the Galactic binary NS systems have a mass less than or equal to $2.7M_\odot$, as listed in Table\ref{tab:mass_data}), the merger remnant should be SMNSs as long as $j/j_{\rm kep}\geq 0.8$ and $M_{\rm max}\geq 2.2M_\odot$ (note that $j\geq 0.8j_{\rm kep}$ unless the postmerger gravitational wave radiation is more efficient than that found in the literature \cite{Zappa2018,Radice2018,Shao2020,Fan2020}), in agreement with previous research (e.g., Refs. \citep{Morrison2004, Lawrence2015, Piro2017, Ma2018}).

\begin{figure}[!ht]
\begin{center}
\includegraphics[width=0.7\textwidth]{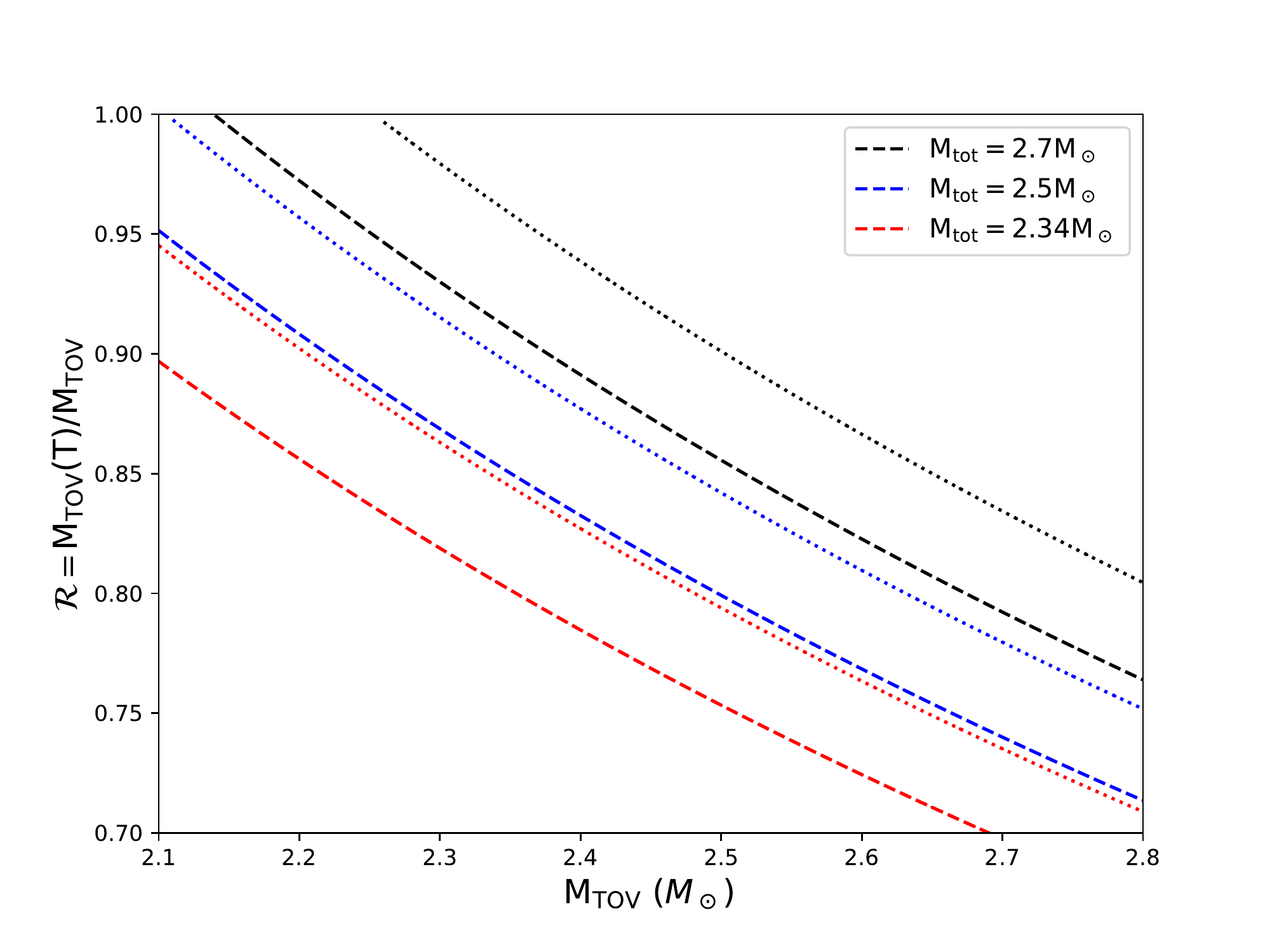}
\end{center}
\caption{The ``hypothesized" ${\cal R}$ to collapse for the given $M_{\rm tot}$ (in particular for the lightest NSs or double NS binaries) and the zero temperature $M_{\rm TOV}$. The dashed and dotted lines are for $j/j_{\rm kep}=1$ and $0.8$, respectively. A minimum $M_{\rm tot}=2.34M_\odot$ is adopted because the lightest neutron star detected so far has an accurately measured mass of $1.17M_\odot$ \citep{Martinez2015}. For simplicity, the best-fit $\zeta_{\rm TOV}-M_{\rm TOV}$ relation presented in Fig.\ref{fig:MR_tov} is adopted.}
\label{fig:post-merger}
\end{figure}

There is, however, one caution that Eqs.(\ref{eq:M_tot_c}) and (\ref{eq:M_tot_c_new}) are derived with some empirical relationships that are established for a group of representative EoSs of NSs at zero temperature. For the nascent NSs formed in the mergers, the typical temperature is found to be approximately $30-50$ MeV (e.g., Ref.\citep{Hotokezaka2013a}).
At such high temperatures, strong interactions may sizably enhance the number density of negatively charged pions.
Very recently,  \citet{Fore2020} have calculated such an effect for a range of density (below 1.4 times of the nuclear saturation density) and temperature using the virial expansion and found that the thermal pions increase the proton fraction and soften the EoS. Therefore a new parameter ${\cal R}\equiv M_{\rm TOV}(T)/M_{\rm TOV}\leq 1$  may be introduced to describe the {\it potential} reduction of the ``effective" maximum gravitational mass at the high temperature ($T$ is approximately tens MeV; note that if it finally turns out that ${\cal R}>1$ the prospect of forming SMNSs in the double NS mergers would be even more promising than estimated in this work). In reality, ${\cal R}$ may be a complicated function. For illustration we simply assume a constant and then discuss the range of ${\cal R}$ that can be probed in the near future. As demonstrated in Fig.\ref{fig:post-merger}, in principle, the mergers of some light double NS binary systems can effectively probe ${\cal R}$ in a wide range supposing the remnant nature (hypermassive NS, SMNS, or even stable NS) can be reasonably inferred from the follow-up electromagnetic observations. Therefore, the possible effect of high temperature on softening the EoS can be unambiguously clarified.

\subsection{Energy reservoir of SMNSs and the acceleration of EeV cosmic rays}
\begin{figure}[!ht]
\begin{center}
\includegraphics[width=0.45\textwidth]{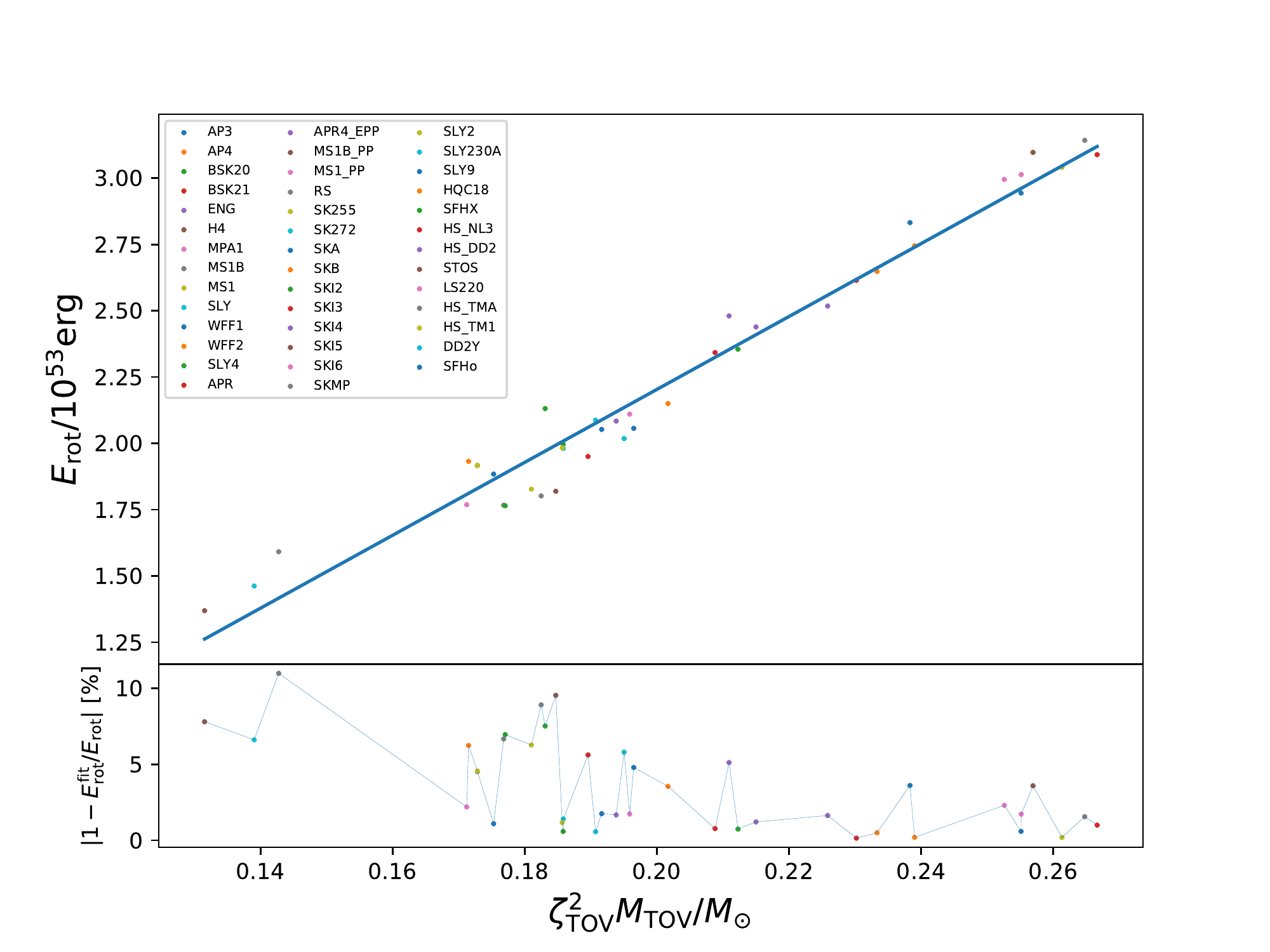}
\includegraphics[width=0.45\textwidth]{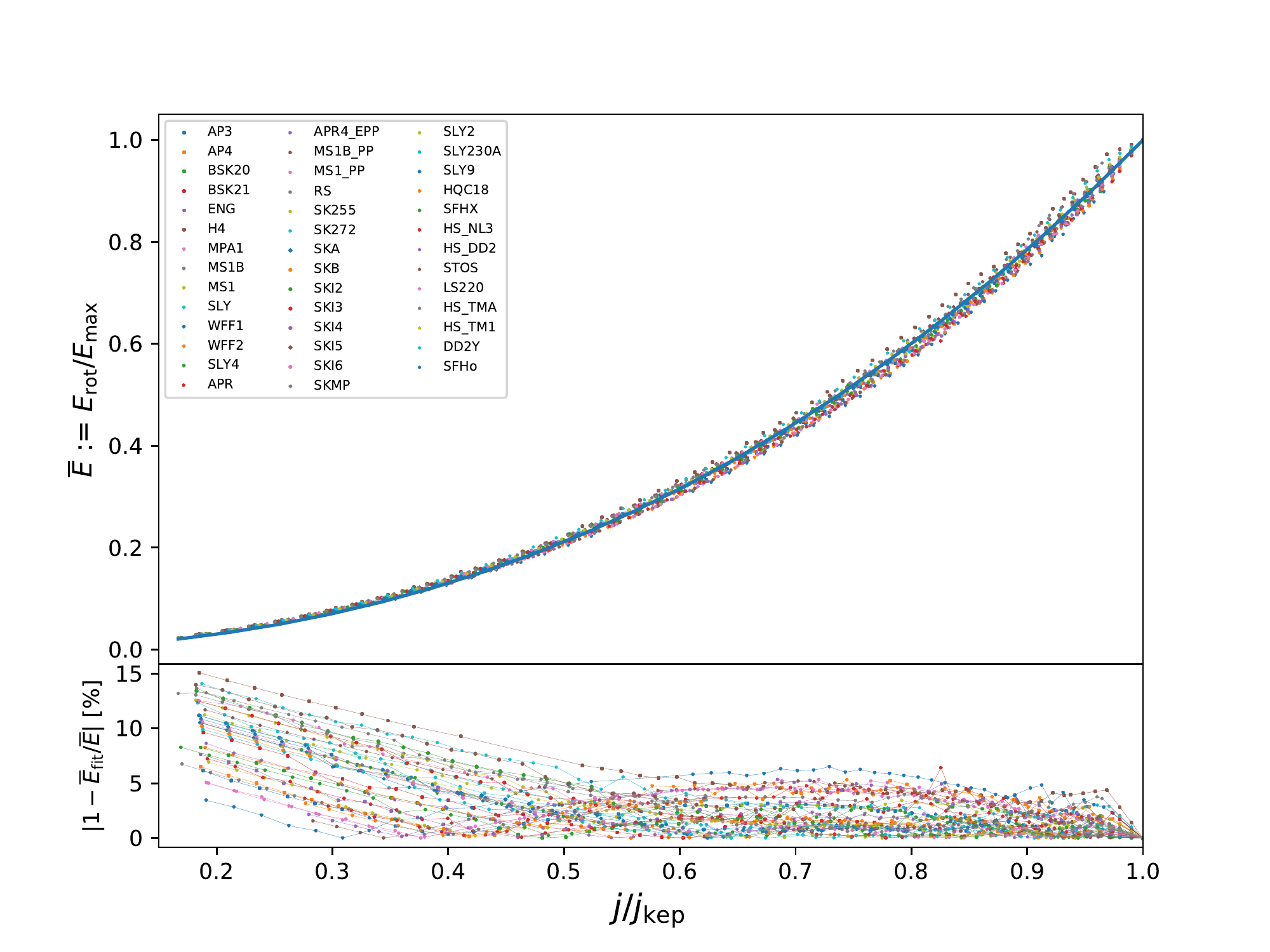}
\end{center}
\caption{Left panel: kinetic rotational energy of NS with mass shedding configuration. Right panel: kinetic rotational energy of NS at the critical point of certain angular momentum.}
\label{fig:E_rot_j}
\end{figure}

The SMNSs, if formed, have a huge amount of energy that could give rise to some very interesting phenomena. In this subsection, we estimate the kinetic rotational energy of these objects and discuss the possible role in accelerating EeV cosmic rays. In the Appendix of \citep{Shao2020}, we have introduced the details of deriving the empirical relation among $M_{\rm crit}$, $M_{\rm TOV}$, and $j/j_{\rm kep}$. Our current approach is basically the same as \citep{Shao2020} except that the EoS sample has been further expanded to include also STOS \citep{1998NuPhA.637..435S}, H4 \citep{2006PhRvD..73b4021L}, MS1B, MS1, MS1B\_PP, MS1\_PP \citep{1996NuPhA.606..508M}, AP3, APR \citep{1971ApJ...170..299B, Akmal1998, 2001A&A...380..151D}, SKI5 \citep{1995NuPhA.584..467R, 2009NuPhA.818...36D, 2015PhRvC..92e5803G}, HS\_TM1, HS\_TMA, HS\_NL3 \citep{Sugahara1994, 1995NuPhA.588..357T, 1997PhRvC..55..540L, 2005PThPh.113..785G, 2010NuPhA.837..210H}. We aim to get an empirical relationship among $E_{\rm rot}$, $M_{\rm TOV}$, $R_{\rm TOV}$ and $j/j_{\rm kep}$. For our purpose, first, we examine the rotational kinetic energy of the SMNS at the mass shedding limit. Motivated by the facts of $E_{\rm rot,kep}\propto I_{\rm kep}\Omega_{\rm kep}^{2}$, $I_{\rm kep}\propto \zeta_{\rm kep}M_{\rm kep}R_{\rm kep}^{2}$ \citep{Koliogiannis2020}, $M_{\rm kep}\propto M_{\rm TOV}$, $R_{\rm kep}\propto R_{\rm TOV}$ and $\Omega_{\rm kep}\propto \sqrt{M_{\rm TOV}/R_{\rm TOV}^{3}}$ \citep{Breu2016}, we have $E_{\rm rot,kep}\propto \zeta_{\rm TOV}^{2}M_{\rm TOV}$. The polynomial fit to the numerical results of a set of widely discussed EOSs yields (see the left panel of Fig.\ref{fig:E_rot_j})
\be
E_{\rm rot,kep}\approx 10^{53}~{\rm erg}~[13.745(\zeta_{\rm TOV}^2M_{\rm TOV}/1M_\odot)-0.546].
\ee

In the right panel of Fig.\ref{fig:E_rot_j}, we show the results of $E_{\rm rot}/E_{\rm rot,kep}$ as a function of $j/j_{\rm kep}$, which reads
\be
E_{\rm rot}(j)=E_{\rm rot,kep}\left[0.69\left({j}/{j_{\rm kep}}\right)^2+0.31\left({j}/{j_{\rm kep}}\right)^3\right],
\label{eq:E_rot_j}
\ee
with which we can estimate the total amount of kinetic energy lost before the collapse of the SMNS if its initial (i.e., at the birth, $j_{\rm int}$) and final (i.e., at the collapse, $j_{\rm c}$) values of the dimensionless angular momentum $j$ are known. The fit for $j/j_{\rm kep}<0.5$ is relatively poor, but it is still acceptable since the kinetic energy is already small in such a range. It is challenging to reliably infer $j_{\rm int}$ with the current information from the observations and the numerical simulations (see Refs.\citep{Shibata2019, Shao2020} for the extended discussion). Anyhow, the dominant contribution of the early time (in the first approximately $100$ms or so) angular momentum loss is likely due to the postmerger gravitational wave radiation. If such a process just carries away the energy of a few percent of solar mass (i.e., in the low end found in Ref.\citep{Bernuzzi2016}), one would have $j_{\rm int}\approx j_{\rm kep}$ (see e.g., Refs.\citep{Piro2017, Zappa2018}).

{}

After the formation of SMNS (or stable NS), the quadrupole gravitational wave radiation is approximately $10^{41} (\epsilon/10^{-8})^{2}(P_{\rm int}/1{\rm ms})^{-6}~{\rm erg~s^{-1}}$ (note that the upper limit on the ellipticity set by LIGO/Virgio observations for the Galactic pulsars is $\epsilon \leq 10^{-7}$\citep{LIGO-NS2019}. Then, usually the energy release would be dominated by magnetic dipole radiation and the spindown luminosity can be estimated \citep{Shapiro1983} by
\begin{equation}
L_{\rm dip}\approx 10^{47} ~\rm erg~s^{-1}~\left(\frac{R_{\rm s}}{10^6~\rm cm}\right)^6\left(\frac{B_\perp}{10^{14} \rm G}\right)^2\left(\frac{P_{\rm int}}{1 \rm ms}\right)^{-4},
\end{equation}
where $R_{\rm s}$ is the radius of SMNS, $B_\perp=B_{\rm s}~\sin{\alpha}$, $P_{\rm int}$ is the initial spin period of the SMNS, and $B_{\rm s}$ is the magnetic field on the surface and $\alpha$ is the angle between the spin and dipole axes. Then the dipole radiation energy injected into the blast wave and ejecta is  $dE_{\rm inj}/dt\approx L_{\rm dip}(1+t/\tau)^{-2}$, where $\tau=E_{\rm rot,int}/L_{\rm dip}$ is the so-called spindown timescale. For quickly rotating SMNS, supposing $R_{\rm s}\sim 14 {\rm km}$ and $P_{\rm int}\sim0.7 {\rm ms}$, we have $\tau \approx 3\times 10^{4}~{\rm s}~(E_{\rm rot,int}/10^{53}~{\rm erg})(R_{\rm s}/14~{\rm km})^{-6}(B_\perp/10^{14}~{\rm G})^{-2}(P_{\rm int}/0.7{\rm ms})^4$. Please bear in mind that at $t\approx \tau$, the total energy released is $E_{\rm inj}\approx 3E_{\rm rot,int}/4$, before which $j$ may have already dropped below $j_{\rm c}$ and the SMNS has collapsed. In this work, for illustration, we ignore such a possibility and simply estimate the possible EeV cosmic ray accelerated in the SMNS wind-driving blast wave (see also Ref.\cite{Li2014}, in which the discussions, however, are mixed with other possible models of the fast radio bursts and are hence not systematic). If the frequency $\omega_{\rm wind}=2/P_{\rm int}\approx 3000~{\rm Hz}~(P_{\rm int}/0.7~{\rm ms})^{-1}$ of the electromagnetic wind of the SMNS is lower than the plasma frequency $\omega_{\rm plasma}\approx 5.6\times 10^4~{\rm Hz}~n_{\rm e}^{1/2}$, the wind would be absorbed by the surrounding material (either the shocked circumburst medium or the merger-driving subrelativistic outflow). Such a condition can be rewritten as $n_{\rm e,c}\geq 3\times 10^{-3}~{\rm cm^{-3}}$.

\begin{figure}[!ht]
\begin{center}
\includegraphics[width=0.7\textwidth]{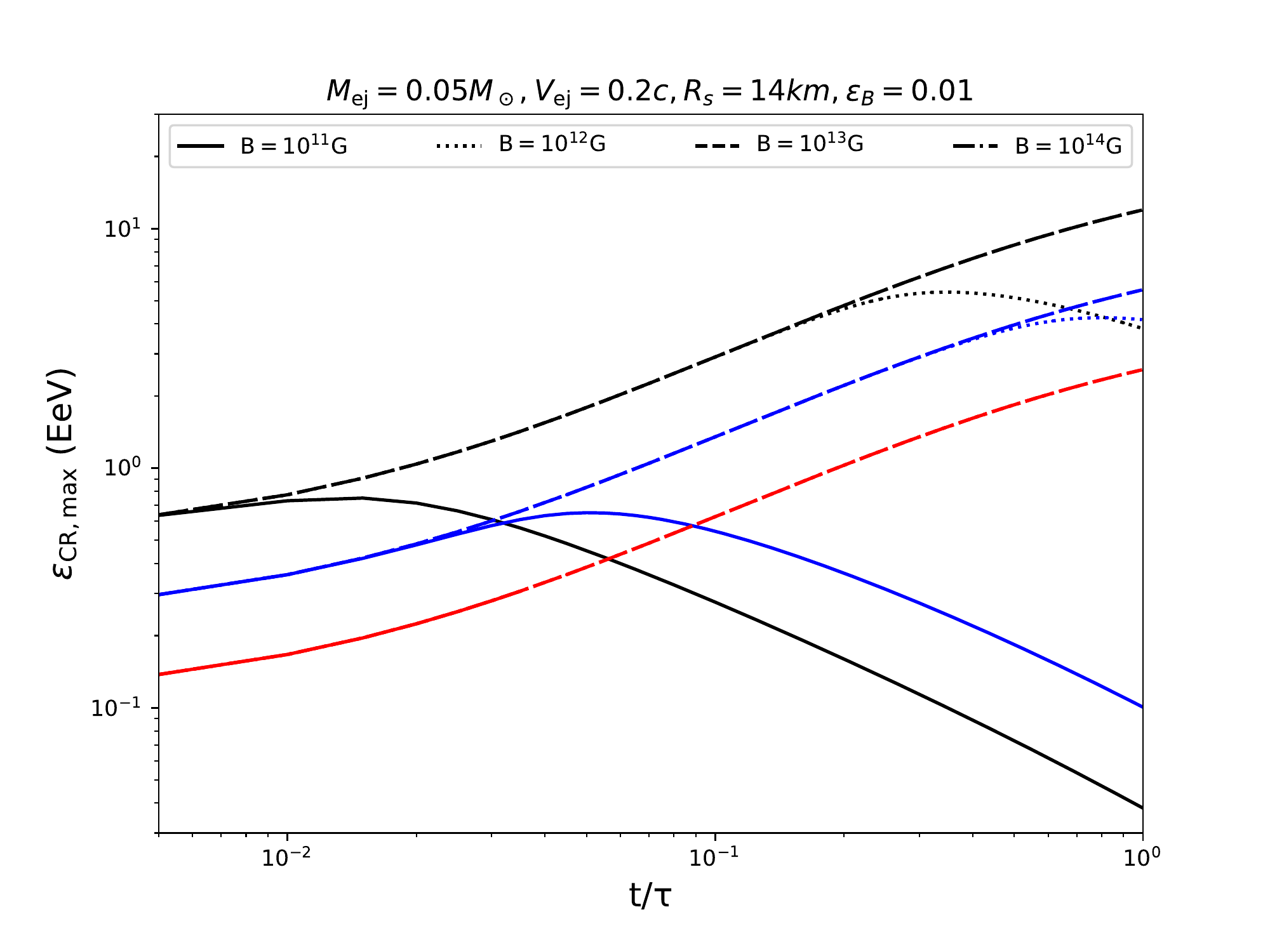}
\end{center}
\caption{Capability of the shocks driven by the kilonova ejecta and the possible energy injection from the central SMNS. The solid, dotted, dashed, and dot-dashed lines are for $B_{\rm s}= 10^{11}, 10^{12}, 10^{13}, 10^{14}$ G, respectively, while the colors of the lines (black, blue, red) correspond to the cases of $n_{\rm m} = (10^{-1}, 10^{-3}, 10^{-5})~{\rm cm^{-3}}$ respectively. For $n_{\rm m} = 10^{-5}~{\rm cm^{-3}}$, the red solid line terminates at $t/\tau\sim0.03$ because at later times we have $R>1.7\times 10^{19}$cm, for which the plasma frequency is lower than the frequency of pulsar wind and the pulsar wind will escape freely rather than be absorbed by the outmoving ejecta.}
\label{fig:eps_max}
\end{figure}

If the absorption is dominated by the merger-driving subrelativistic outflow (i.e., the circumburst medium has a low number density $n_{\rm m}\ll n_{\rm e,c}$) with a mass of $M_{\rm ej}\sim 0.05M_\odot$, we would have $R_{\rm c}\leq 1.7\times 10^{19}~{\rm cm}~(M_{\rm ej}/0.05M_\odot)^{1/3}$ and the bulk Lorentz factor of the outflow will peak with $\Gamma_{\tau}\approx 1+(E_{\rm inj}/10^{53}~{\rm erg})(M_{\rm ej}/0.05M_\odot)^{-1}$, supposing $\tau< R_{\rm c}/c \sim 5.6\times 10^{8}~{\rm s}~(M_{\rm ej}/0.05M_\odot)^{1/3}$ (i.e., $B_\perp\geq 10^{12}~{\rm G}$). The deceleration radius can be estimated to be $R_{\rm dec}\sim (M_{\rm ej}/\Gamma_{\tau}n_{\rm m}m_{\rm p})^{1/3}\sim 4.2\times 10^{19}~{\rm cm}~(M_{\rm ej}/0.05M_\odot)^{1/3}(\Gamma_{\tau}/2)^{-1/3}(n_{\rm m}/10^{-4}~{\rm cm^{-3}})^{-1/3}$, where $m_{\rm p}$ is the rest mass of proton. The maximum energy of the accelerated particles can be estimated as
\be
\varepsilon_{\rm CR,max}\sim \beta ZeB_{\rm dec}R_{\rm dec} \sim 4\times 10^{18}~{\rm eV}~Z\beta^2 \Gamma\left(\frac{n_{\rm m}}{10^{-4}~{\rm cm^{-3}}}\right)^{\frac{1}{6}}\left(\frac{M_{\rm ej}}{0.05~M_\odot}\right)^{\frac{1}{3}}\left(\frac{\epsilon_{\rm B}}{10^{-2}}\right)^{\frac{1}{2}},
\ee
where $B_{\rm dec}\sim 3.2\times 10^{-4}~{\rm G}~\beta\Gamma(n_{\rm m}/10^{-4}~{\rm cm^{-3}})^{1/2}(\epsilon_{\rm B}/0.01)^{1/2}$ is the strength of shock amplified magnetic field at $R_{\rm dec}$. In the case of the binary NS mergers taking place in the relatively dense circumburst medium (i.e., $n_{\rm m}\sim 0.1~{\rm cm^{-3}}$), the deceleration of the blast wave starts at the radius of $R_{\rm dec}\sim (M_{\rm ej}/\Gamma_{\tau}n_{\rm m}m_{\rm p})^{1/3}\sim 4.2\times 10^{18}~{\rm cm}~(M_{\rm ej}/0.05M_\odot)^{1/3}(\Gamma_{\tau}/2)^{-1/3}(n_{\rm m}/0.1~{\rm cm^{-3}})^{-1/3}$, and most of the kinetic energy of the SMNS would have been injected into the blast wave supposing $B_\perp\geq 4\times 10^{12}~{\rm G}$. In this case, efficient EeV protons are still accelerated. For $B_\perp \ll 4\times 10^{12}~{\rm G}$, the significant energy injection lasts much longer time and the accelerated protons can only reach sub-EeV energy region. To widely explore the various possibilities, we present the numerical calculation results in Fig.\ref{fig:eps_max}. Clearly, for $B_\perp>10^{12}$ G, EeV cosmic-ray protons are indeed the plausible outputs.

The EeV cosmic-ray proton flux accelerated by the double NS merger formed SMNSs at approximately $10^{18}$ eV is estimated as
\be
F_{\rm EeV-CR} \sim 10^{-28}\eta\left(\frac{\omega}{0.1}\right)\left(\frac{E_{\rm inj}}{10^{53}~{\rm erg}}\right)\times\left(\frac{\cal R_{\rm DNS,SMNS}}{10^3 {\rm yr^{-1}Gpc^{-3}}}\right)~{\rm m^{-2}s^{-1}sr^{-1}eV^{-1}},
\ee
which is consistent with the observed cosmic-ray flux $F_{\rm obs}(\varepsilon_{\rm CR})=C(\varepsilon_{\rm CR}/6.3\times10^{18}~{\rm eV})^{-3.2\pm0.05}$ with $C=(9.23\pm0.065)\times10^{-33}~{\rm m^{-2}s^{-1}sr^{-1}eV^{-1}}$ \citep{Nagano2000}, for the EeV cosmic-ray acceleration efficiency $\eta\sim 0.03$ and $\omega\sim 0.1$, the fraction of total cosmic-ray energy at each energy decade. Here we normalize the SMNS formation rate to that of the double NS mergers \citep{GW170817} because SMNSs may be produced in a good fraction of such a kind of gravitational wave events (see Fig.\ref{fig:Mtotc}). A good fraction of double NS mergers is expected to take place in the elliptical galaxies that are short of the star formation now \citep{Berger2014}. If some EeV cosmic rays or PeV($10^{15}$eV) neutrinos can be found in the directions of the elliptical galaxies, our arguments will be strongly favored.

\section{Conclusions}
The mass distribution of the neutron stars is known to be important in shedding light on the supernova explosion and the accretion dynamics of binary neutron star. Moreover, the inferred cutoff mass may represent the maximum mass of the nonrotating cold neutron stars, which is essential for revealing the equation of state of matter with ultrahigh density. However, the mass measurements are challenging and the sample just consists of 74 objects in the work by \citet{Alsing2018}. Thanks to the dedicated worldwide joint efforts, recently, the sample of neutron stars with a measured mass has been growing very quickly. In this work, we collect the mass measurements of NSs from the latest literature (including the updates of the masses of some ``old" objects) and increase the total number of neutron stars in the sample to 103. With this new sample, we adopt a flexible two-component mixture model and a Gaussian plus Cauchy-Lorentz component model to infer the mass distribution of neutron stars and use Bayesian model selection to explore evidence for multimodality and a sharp cutoff in the mass distribution. The results for these two models are consistent with each other (see Fig.\ref{fig:mmax}). In agreement with previous studies, we find evidence for a bimodal distribution together with a cutoff at a mass of $M_{\rm max}=2.26_{-0.05}^{+0.12}M_\odot$ (68\% credible interval; for the 95\% credible interval, $M_{\rm max}=2.26_{-0.11}^{+0.47}M_\odot$). Our $M_{\rm max}$ is larger than that reported by \citet{Alsing2018} by approximately $0.1M_\odot$ mainly because of the inclusion of the recent measurements of two massive objects PSR J1959+2048 and PSR J2215+5135 \citep{Kandel2020} (for PSR J2215+5135 there was an independent mass measurement of $2.27^{+0.17}_{-0.15}M_\odot$ \citep{Linares2018}, with which we have $M_{\rm max}=2.22^{+0.10}_{-0.06}M_\odot$). Compared to previous works, our method faithfully reproduces the asymmetric mass measurement errors and avoids the bias caused by the approximation of Gaussian measurement errors. Our resulting distributions of $\mu_1$, $\sigma_1$, and $M_{\rm max}$ are less affected by the priors and NS population models, indicating the robustness of the approaches. We have  discussed some possible selection efforts, for example that our sample mainly consists of the neutron stars found in binary systems and the most accurately measured objects are from the double neutron star systems. However, the most massive neutron stars are heavily recycled and the isolated neutron stars are not expected to have a mass close to $M_{\rm max}$. We also show that both PSR J0348+0432 and PSR J2215+5135, the two objects play the major role in bounding $M_{\rm max}$, are among the luminous radio pulsars. Hence at least for the current sample, there is no evidence for a sizeable nondetecion/misidentification probability of the very massive NSs in radio (these recycled pulsars rotate very quickly, which generate strong radio radiation though the surface magnetic fields are low). In agreement with \citet{Alsing2018}, we suggest that the selection effects are unlikely to account for the inferred hard cutoff at $M_{\rm max}$. We further examined the possible dependence between radio luminosity and mass but did not find any evidence. Together with the previous findings \citep{Lorimer2006,Kiziltan2013}, we suggest that the selection effects will not introduce serious bias to the observed mass distribution (see also Ref.\citep{Kiziltan2013}).

If the inferred cutoff mass $M_{\rm max}$ represents the maximum gravitational mass of nonrotating cold neutron stars ($M_{\rm TOV}$), the prospect of forming supramassive remnants is found to be promising for the double neutron star mergers with a total gravitational mass less than or equal to $2.7M_\odot$ (see Fig.\ref{fig:Mtotc}) unless the thermal pions could substantially soften the equation of state for the very hot neutron star matter or alternatively the postmerger gravitational wave radiation has carried away the kinetic rotational energy of approximately $0.1M_\odot$ (for which $j/j_{\rm kep}<0.8$). As demonstrated in Fig.\ref{fig:post-merger}, the mergers of the light double NS binary systems, such as PSR J0514-4002A ($M_{\rm tot}=2.47M_\odot$) and PSR J1946+2052 ($M_{\rm tot}=2.50M_\odot$), may effectively probe the potential effect of EoS softening by the thermal pions generated at high temperatures, supposing the remnant nature (hypermassive NS, SMNS or even stable NS) can be reliably inferred from the follow-up electromagnetic observations. The SMNSs are expected to have a typical kinetic rotational energy of approximately $1-2\times 10^{53}$ ergs. If not radiated mainly in a gravitational wave, thanks to a high neutron star merger rate of approximately $10^{3}~{\rm Gpc^{-3}~yr^{-3}}$ and the plausible high chance of forming SMNSs in such mergers, the neutron star mergers are likely the significant sources of EeV cosmic-ray protons.

In the O3 run of advanced LIGO/Virgo, almost all double neutron star event/candidates were just detected by one of the two LIGO detectors (https://gracedb.ligo.org/superevents/public/O3/). These events were poorly localized and no electromagnetic counterparts were reliably identified. The situation will change considerably in the late observing runs (see https://dcc.ligo.org/public/0161/P1900218/002/SummaryForObservers.pdf for the latest schedule for the future plans of the second generation gravitational wave detectors). The Kamioka Gravitational Wave Detector (KAGRA) had already joined the O3 run in March 2020. Moreover, the sensitivities of Virgo and KAGRA detectors will be enhanced by a factor of a few in the O4 run that is expected to start in January 2022. LIGO-India is anticipated to join the O5 run in 2025. So the future gravitational wave events will be significantly better localized and their electromagnetic counterparts, in particular the kilonovae/macronovae, will be much more frequently detected. These events are expected to be able to test some speculations of this work, in particular the possibilities of ${\cal R}<1$, and the formation of SMNSs in a good fraction of neutron star mergers.

\section*{Acknowledgments}
We thank the referees and Professor C. M. Zhang for very helpful suggestions. This work was supported in part by NSFC under Grants No. 11525313 (i.e., Funds for Distinguished Young Scholars) and No. 11921003.

\appendix
\section{Neutron star data}
Here, we summarize the measurements of the neutron star masses, totally 103 items classified into six types. This sample is updated to February 2020.

\begin{longtable}{lcccccl}
\caption{Catalog of neutron stars with mass measurements \label{tab:mass_data}}\\
\toprule
\hline
{Name} & {Type} & {$m_{\rm p}$ ($M_\odot$)} & {$f$ ($M_\odot$)} & {$m_{\rm T}$ ($M_\odot$)} & {$q$} & {Reference}\\
\hline
\midrule
\endfirsthead

\multicolumn{7}{c}{\autoref{tab:mass_data} ({\it Continued})}\\
\toprule
\hline
{Name} & {Type} & {$m_{\rm p}$ [$M_\odot$]} & {$f$ [$M_\odot$]} & {$m_{\rm T}$ [$M_\odot$]} & {$q$} & {Reference}\\
\hline
\midrule
\endhead
\bottomrule

\hline\\
\multicolumn{7}{c}{\autoref{tab:mass_data} ({\it Continued on next page})}\\
\endfoot
\hline\\
\multicolumn{7}{l}{
Notes: NS-NS, double neutron star system; NS-WD, neutron star-white dwarf binary; NS-MS, neutron star-main sequence star system;}\\
\multicolumn{7}{l}{
HMXB, high mass x-ray binary; LMXB, low mass x-ray binary; INS, isolated neutron star.}\\
\multicolumn{7}{l}{
The question mark means the nature of the companion is uncertain.}\\
\endlastfoot

B1534+12 & NS-NS & 1.3330$\pm$0.0002 & & & & \citet{Fonseca2014}\\
B1534+12 comp. & NS-NS & 1.3455$\pm$0.0002 & & & & \citet{Fonseca2014}\\
B1913+16 & NS-NS & 1.4398$\pm$0.0002 & & & & \citet{Weisberg2010}\\
B1913+16 comp. & NS-NS & 1.3886$\pm$0.0002 & & & & \citet{Weisberg2010}\\
B2127+11C & NS-NS & 1.358$\pm$0.010 & & & & \citet{Jacoby2006}\\
B2127+11C comp. & NS-NS & 1.354$\pm$0.010 & & & & \citet{Jacoby2006}\\
J0453+1559 & NS-NS & 1.559$\pm$0.004 & & & & \citet{Martinez2015}\\
J0453+1559 comp. & NS-NS & 1.174$\pm$0.004 & & & & \citet{Martinez2015}\\
J0509+3801 & NS-NS & 1.34$\pm$0.08 & & & & \citet{Lynch2018}\\
J0509+3801 comp. & NS-NS & 1.46$\pm$0.08 & & & & \citet{Lynch2018}\\
J0514-4002A & NS-NS & $1.25^{+0.05}_{-0.06}$ & & & & \citet{Ridolfi2019}\\
J0514-4002A comp.& NS-NS & $1.22^{+0.06}_{-0.05}$ & & & & \citet{Ridolfi2019}\\
J0737-3039A & NS-NS & 1.3381$\pm$0.0007 & & & & \citet{Kramer2006}\\
J0737-3039B & NS-NS & 1.2489$\pm$0.0007 & & & & \citet{Kramer2006}\\
J1756-2251 & NS-NS & 1.341$\pm$0.007 & & & & \citet{Ferdman2014}\\
J1756-2251 comp. & NS-NS & 1.230$\pm$0.007 & & & & \citet{Ferdman2014}\\
J1757-1854 & NS-NS & 1.3384$\pm$0.0009 & & & & \citet{Cameron2018}\\
J1757-1854 comp. & NS-NS & 1.3946$\pm$0.0009 & & & & \citet{Cameron2018}\\
J1807-2500B & NS-NS & 1.3655$\pm$0.0021 & & & & \citet{Lynch2012}\\
J1807-2500B comp. & NS-NS & 1.2064$\pm$0.0020 & & & & \citet{Lynch2012}\\
J1906+0746 & NS-NS & 1.291$\pm$0.011 & & & & \citet{vanLeeuwen2014}\\
J1906+0746 comp. & NS-NS & 1.322$\pm$0.011 & & & & \citet{vanLeeuwen2014}\\
GW170817A & NS-NS & $1.47^{+0.09}_{-0.07}$ & & & & \citet{GW170817}\\
GW170817B & NS-NS & $1.27^{+0.06}_{-0.07}$ & & & & \citet{GW170817}\\
GW190425A & NS-NS & $1.56^{+0.06}_{-0.08}$ & & & & \citet{GW190425}\\
GW190425B & NS-NS & $1.74^{+0.10}_{-0.06}$ & & & & \citet{GW190425}\\
J1411+2551 & NS-NS & & 0.1223898 & 2.538$\pm$0.022 & & \citet{Martinez2017}\\
J1518+4904 & NS-NS & & 0.115988 & 2.7183$\pm$0.0007 & & \citet{Janssen2008}\\
J1811-1736 & NS-NS & & 0.128121 & 2.57$\pm$0.10 & & \citet{Corongiu2007}\\
J1829+2456 & NS-NS & & 0.29413 & 2.59$\pm$0.02 & & \citet{Champion2005}\\
J1913+1102 & NS-NS & & 0.136344 & 2.875$\pm$0.014 & &\citet{Lazarus2016}\\
J1930-1852 & NS-NS & & 0.34690765 & 2.54$\pm$0.03 & & \citet{Swiggum2015}\\
J1946+2052 & NS-NS & & 0.268184 & 2.50$\pm$0.04 & & \citet{Stovall2018}\\
B1855+09 & NS-WD & 1.37$\pm$0.13 & & & & \citet{Arzoumanian2018}\\
J0337+1715 & NS-WD & 1.4359$\pm$0.0003 & & & & \citet{Archibald2018}\\
J0348+0432 & NS-WD & 2.01$\pm$0.04 & & & & \citet{Antoniadis2013}\\
J0437-4715 & NS-WD & 1.44$\pm$0.07 & & & & \citet{Reardon2016}\\
J0621+1002 & NS-WD & $1.53^{+0.10}_{-0.20}$ & & & & \citet{Kasian2012}\\
J0740+6620 & NS-WD & $2.14^{+0.10}_{-0.09}$ & & & & \citet{Cromartie2020}\\
J0751+1807 & NS-WD & 1.64$\pm$0.15 & & & & \citet{Desvignes2016}\\
J1012+5307 & NS-WD & 1.83$\pm$0.11 & & & & \citet{Antoniadis2016}\\
J1141-6545 & NS-WD & 1.27$\pm$0.01 & & & & \citet{Krishnan2020}\\
J1600-3053 & NS-WD & $2.3^{+0.7}_{-0.6}$ & & & & \citet{Arzoumanian2018}\\
J1614-2230 & NS-WD & 1.908$\pm$0.016 & & & & \citet{Arzoumanian2018}\\
J1713+0747 & NS-WD & 1.33$\pm$0.10 & & & & \citet{Zhu2019a}\\
J1738+0333 & NS-WD & 1.47$\pm$0.07 & & & & \citet{Antoniadis2012}\\
J1741+1351 & NS-WD & $1.14^{+0.43}_{-0.25}$ & & & & \citet{Arzoumanian2018}\\
J1748-2446am & NS-WD & $1.649^{+0.037}_{-0.11}$ & & & & \citet{Andersen2018}\\
J1802-2124 & NS-WD & 1.24$\pm$0.11 & & & & \citet{Ferdman2010}\\
J1811-2405 & NS-WD & $2.0^{+0.8}_{-0.5}$ & & & & \citet{Ng2020}\\
J1909-3744 & NS-WD & 1.48$\pm$0.03 & & & & \citet{Arzoumanian2018}\\
J1911-5958A & NS-WD & 1.34$\pm$0.08 & & & & \citet{Bassa2006}\\
J1918-0642 & NS-WD & 1.29$\pm$0.1 & & & & \citet{Arzoumanian2018}\\
J1946+3417 & NS-WD & 1.828$\pm$0.022 & & & & \citet{Barr2017}\\
J1949+3106 & NS-WD & $1.34^{+0.17}_{-0.15}$ & & & & \citet{Zhu2019}\\
J1950+2414 & NS-WD & 1.496$\pm$0.023 & & & & \citet{Zhu2019}\\
J1959+2048 & NS-WD & 2.18$\pm$0.09 & & & & \citet{Kandel2020}\\
J2043+1711 & NS-WD & 1.38$\pm$0.13 & & & & \citet{Arzoumanian2018}\\
J2045+3633 & NS-WD & 1.33$\pm$0.3 & & & & \citet{Berezina2017}\\
J2053+4650 & NS-WD & 1.40$\pm$0.21 & & & & \citet{Berezina2017}\\
J2215+5135 & NS-WD & $2.28^{+0.10}_{-0.09}$ & & & & \citet{Kandel2020}\\
J2222-0137 & NS-WD & 1.76$\pm$0.06 & & & & \citet{Cognard2017}\\
J2234+0611 & NS-WD & $1.353^{+0.014}_{-0.017}$ & & & & \citet{Stovall2019}\\
B1516+02B & NS-WD & & 0.000646723 & 2.29$\pm$0.17 & & \citet{Freire2008a}\\
B1802-07 & NS-WD & & 0.00945034 & 1.62$\pm$0.07 & & \citet{Thorsett1999}\\
B2303+46 & NS-WD & & 0.246261924525 & 2.64$\pm$0.5 & & \citet{Thorsett1999}\\
J0024-7204H & NS-WD(?) & & 0.001927 & 1.665$\pm$0.007 & &\citet{Freire2017}\\
J1748-2446I & NS-WD & & 0.003658 & 2.17$\pm$0.02 & & \citet{Ransom2005}\\
J1748-2446J & NS-WD & & 0.013066 & 2.20$\pm$0.04 & & \citet{Ransom2005}\\
J1750-37A & NS-WD & & 0.0518649 & 1.97$\pm$0.15 & & \citet{Freire2008}\\
J1824-2452C & NS-WD & & 0.006553 & 1.616$\pm$0.007 & & \citet{Freire2008a}\\
NGC6440B & NS-WD & & 0.0002266235 & 2.69$\pm$0.071 & & \citet{Clifford2019}\\
J1311-3430 & NS-WD(?) & & $3\times10^{-7}$ & & 175$\pm$3 & \citet{Romani2012}\\
J1723-2837 & NS-WD(?) & & 0.005221 & & 3.45$\pm$0.02 & \citet{vanStaden2016}\\
J1740-5340 & NS-WD(?) & & 0.002644 & & 5.85$\pm$0.13 & \citet{Ferraro2003}\\
J1816+4510 & NS-WD(?) & & 0.0017607 & & 9.54$\pm$0.21 & \citet{Kaplan2013}\\
J0045-7319 & NS-MS & 1.58$\pm$0.34 & & & & \citet{Nice2003}\\
J1023+0038 & NS-MS & 1.71$\pm$0.16 & & & & \citet{Deller2012}\\
J1903+0327 & NS-MS & 1.666$\pm$0.01 & & & & \citet{Arzoumanian2018}\\
J0030+0451 & INS & $1.34^{+0.15}_{-0.16}$ & & & & \citet{Riley2019}(NICER)\\
4U1538-522 & HMXB & 1.02$\pm$0.17 & & & & \citet{Falanga2015}\\
4U1700-377 & HMXB & 1.96$\pm$0.19 & & & & \citet{Falanga2015}\\
Cen X-3 & HMXB & 1.57$\pm$0.16 & & & & \citet{Falanga2015}\\
EXO 1722-363 & HMXB & 1.91$\pm$0.45 & & & & \citet{Falanga2015}\\
Her X-1 & HMXB & 1.07$\pm$0.36 & & & & \citet{Rawls2011}\\
J013236.7+303228 & HMXB & 2.0$\pm$0.4 & & & & \citet{Bhalerao2012}\\
LMC X-4 & HMXB & 1.57$\pm$0.11 & & & & \citet{Falanga2015}\\
OAO 1657-415 & HMXB & 1.74$\pm$0.3 & & & & \citet{Falanga2015}\\
SAX J1802.7-2017 & HMXB & 1.57$\pm$0.25 & & & & \citet{Falanga2015}\\
SMC X-1 & HMXB & 1.21$\pm$0.12 & & & & \citet{Falanga2015}\\
Vela X-1 & HMXB & 2.12$\pm$0.16 & & & & \citet{Falanga2015}\\
XTE J1855-026 & HMXB & 1.41$\pm$0.24 & & & & \citet{Falanga2015}\\
2S 0921-630 & LMXB & 1.44$\pm$0.1 & & & & \citet{Steeghs2007}\\
4U 1608-52 & LMXB & $1.57^{+0.30}_{-0.29}$ & & & & \citet{Ozel2016a}\\
4U1702-429 & LMXB & 1.9$\pm$0.3 & & & & \citet{Nattila2017}\\
4U 1724-207 & LMXB & $1.81^{+0.25}_{-0.37}$ & & & & \citet{Ozel2016a}\\
4U 1820-30 & LMXB & $1.77^{+0.25}_{-0.28}$ & & & & \citet{Ozel2016a}\\
Cyg X-2 & LMXB & 1.71$\pm$0.21 & & & & \citet{Casares2009}\\
KS 1731-260 & LMXB & $1.61^{+0.35}_{-0.37}$ & & & & \citet{Ozel2016a}\\
EXO 1745-248 & LMXB & $1.65^{+0.21}_{-0.31}$ & & & & \citet{Ozel2016a}\\
SAX J1748.9-2021 & LMXB & $1.81^{+0.25}_{-0.37}$ & & & & \citet{Ozel2016a}\\
X 1822-371 & LMXB & 1.96$\pm$0.36 & & & & \citet{MunozDarias2005}\\
XTE J2123-058 & LMXB & 1.53$\pm$0.42 & & & & \citet{Gelino2002}\\
\end{longtable}

\clearpage

\end{document}